\documentclass[preprint]{iucr}              

     \journalcode{}              

\RequirePackage{graphicx}
\usepackage{verbatim,xcolor,soul,relsize}
\usepackage{amsmath,lineno,amssymb,amsthm,mathtools,url}

\theoremstyle{definition}
\newtheorem{dfn}{Definition}[section]

\newtheorem{thm}[dfn]{Theorem}
\newtheorem{prop}[dfn]{Proposition}

\newtheorem{exa}[dfn]{Example}

\newcommand{\includefigure}[1]{#1}
\newcommand{\includeallfigures}[1]{}

\newcommand{\R}{\mathbb R}

\newcommand{\al}{\alpha}

\newcommand{\La}{\Lambda}
\newcommand{\si}{\sigma}
\newcommand{\ep}{\varepsilon}
\newcommand{\ph}{\phi}

\newcommand{\MT}{\mathrm{MT}}
\newcommand{\QI}{\mathrm{QI}}

\newcommand{\sign}{\mathrm{sign}}

\newcommand{\QS}{\mathrm{QS}}

\newcommand{\HS}{\mathrm{HS}}
\newcommand{\RI}{\mathrm{RI}}

\newcommand{\PI}{\mathrm{PI}}

\newcommand{\TC}{\mathrm{TC}}
\newcommand{\TP}{\mathrm{TP}}
\newcommand{\QT}{\mathrm{QT}}

\newcommand{\EP}{\mathrm{EP}}
\newcommand{\SM}{\mathrm{SM}}

\newcommand{\RIS}{\mathrm{RIS}}

\newcommand{\LIS}{\mathrm{LIS}}

\newcommand{\bs}{\hfill $\blacksquare$}
\newcommand{\lra}{\leftrightarrow}

\newcommand{\bd}{\partial}

\newcommand{\matfour}[4]{\left(\begin{array}{ccc}
#1 & #2 \\ #3 & #4 \end{array}\right)}

\newcommand{\ora}[1]{\overrightarrow{#1}}

\begin{document}                  



\title{Geographic-style maps for 2-dimensional lattices}
\shorttitle{Geographic-style maps for 2-dimensional lattices}


\author[a]{Matthew}{Bright}{}{}
\author[a]{Andrew I}{Cooper}{}{}
\cauthor[a]{Vitaliy}{Kurlin}{vitaliy.kurlin@gmail.com}{}

\aff[a]{Materials Innovation Factory, University of Liverpool, \country{UK}}






\keyword{Lattice, obtuse superbase, isometry, invariant, similarity, metric, continuity, Lattice Isometry Space}



\maketitle                        

\begin{synopsis}
The complete invariant-based maps of 2-dimensional lattices reveal continuous distributions for hundreds of thousands of known crystals from the Cambridge Structural Database.
\end{synopsis}

\begin{abstract}
This paper develops geographic-style maps containing 2D lattices in all known crystals parameterised by recent complete invariants.
Motivated by rigid crystal structures, lattices are considered up to rigid motion and uniform scaling.
The resulting space of 2D lattices is a square with identified edges or a sphere without one point.
The new continuous maps show all Bravais classes as low-dimensional subspaces, visualise hundreds of thousands of real crystal lattices from the Cambridge Structural Database, and motivate the development of continuous and invariant-based crystallography.
\end{abstract}

\section{Practical motivations for the problem to continuously classify lattices}
\label{sec:intro}

This paper for crystallographers presents practical applications of 
the companion paper \cite{kurlin2022mathematics} written 
for mathematicians and computer scientists.
\smallskip

A lattice can be considered as a periodic crystal whose atomic motif consists of a single point.
In Euclidean space $\R^n$, a \emph{lattice} $\La\subset\R^n$ consists of all integer linear combinations of basis vectors $v_1,\dots,v_n$, which span a primitive \emph{unit cell} $U$ of $\La$.
\smallskip

Crystallography traditionally splits crystals into finitely many classes, for instance by their space-group types. 
These discrete symmetry-based classifications were suitable for distinguishing highly symmetric crystals and categorising small datasets.
\smallskip
 
Nowadays crystals are simulated and synthesised on an industrial scale.
The Cambridge Structural Database (CSD) contains more than 1.17M existing crystals \cite{groom2016cambridge}. 
Crystal Structure Prediction (CSP) tools generate millions of crystals even for a fixed chemical composition \cite{pulido2017functional}, mostly with P1 symmetry.
This amount of data requires new approaches to mapping huge crystal datasets.
\smallskip

A more important reason for a continuous approach to the similarity of crystals is the inevitability of noise in data.
Thermal vibrations of atoms can give rise to slightly different X-ray patterns under different crystallisation conditions, resulting in similar but distinct structures.
A common solution is to ignore deviations of lattice parameters or atomic coordinates up to a certain threshold, but this simply moves the problem to a different value - 
if we ignore differences of angles and lengths up to $20^\circ$ and $0.1\AA$, 
the boundary between similarity and dissimilarity remains sharp at other values.
\smallskip

Furthermore, we can argue that the use of threshold values
leads to trivial classifications as follows.
Any decision on similarity (denoted by $\sim$) gives rise to a justified classification only if we use \emph{equivalence} relation, which satisfies three axioms: 
\smallskip

\noindent
(1) reflexivity :  any lattice $\La$ is equivalent to itself, so $\La\sim\La$;
\smallskip

\noindent
(2) symmetry :  if $\La\sim\La'$ then $\La'\sim\La$;
\smallskip

\noindent
(3) transitivity :  if $\La\sim\La'$ and $\La'\sim\La''$ then $\La\sim\La''$.
\smallskip

The transitivity axiom splits lattices into well-defined and disjoint \emph{equivalence classes}: the class $[\La]$ consists of all lattices equivalent to $\La$, since if $\La$ is equivalent to $\La'$, which is equivalent to $\La''$, all three lattices are in the same class. 
Previous similarities in \cite{lima1990nomenclature} use numerical thresholds to determine a lattice class, but as Fig.~\ref{fig:lattice_deformations4} illustrates, all lattices can be made equivalent given some threshold. 

\newcommand{\figureone}{
\begin{figure}
\label{fig:lattice_deformations4}
\caption{All lattices continuously deform into each other if we allow any small changes.}
\includegraphics[width=1.0\textwidth]{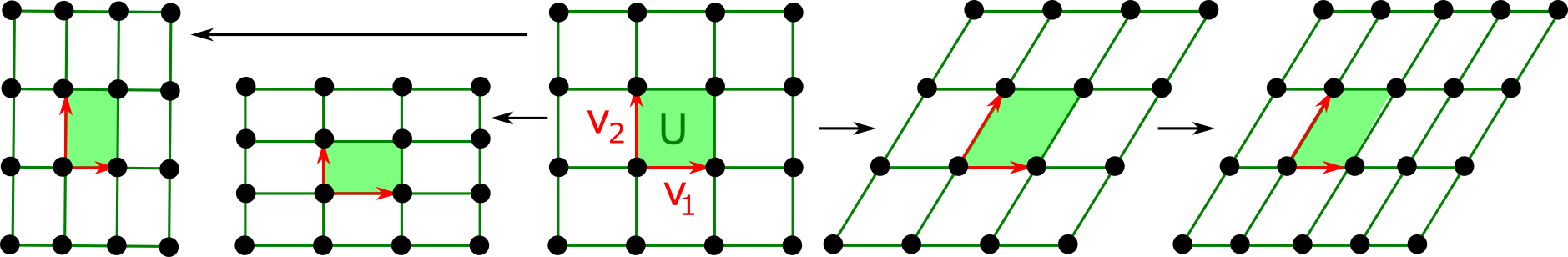}
\end{figure}}
\includefigure{\figureone}

An alternative mathematical approach classifies lattices by space groups and other algebraic structures \cite{nespolo2008does}. 
Since crystal structures are determined as rigid forms, the most practically important equivalence of crystals and their lattices is a \emph{rigid motion}, which in $\R^2$, is any composition of translations and rotations.
This is the strongest possible equivalence on crystals that are indistinguishable as rigid bodies.
\smallskip

Slightly weaker is equivalence based on \emph{isometry} or congruence $\La\cong\La'$ (including reflections), or \emph{similarity} (including uniform scaling).
Even if we fix an equivalence such as isometry, \cite{sacchi2020same} highlights that the key question `same or different' remained unanswered.
What is needed is the notion of an \emph{invariant} $I$ that is a descriptor, such as a numerical vector, taking the same value on all isometric lattices.
In a fixed coordinate system, the basis vectors themselves are not isometry invariants as they easily change under rotation but the primitive cell area does not. 
Crucially, any isometry invariant $I$ has no \emph{false negatives}:  if $\La\cong\La'$ then $I(\La)=I(\La')$. 
Hence, if $I$ takes different values on latices $\La,\La'$, these lattices are certainly not isometric.   
\smallskip

Non-invariants cannot help distinguish equivalent objects.
For example, isometric lattices $\La\cong\La'$ can have infinitely many different bases.
Most isometry invariants allow \emph{false positives} that are non-isometric lattices $\La\not\cong\La'$ with $I(\La)=I(\La')$.
For example, it is easy to imagine an infinite number of non-isometric lattices with the same primitive cell area. 
An isometry invariant $I$ giving rise to no false positives such that $\La\cong\La'$ if and only if $I(\La)=I(\La')$, is called a \emph{complete} (or injective) invariant.
\smallskip

Complete invariants are the main goal of all classifications.
For crystals, we need even better invariants: they must also be continuous under perturbation of a lattice basis, since a discontinuous invariant may take very different values on nearly identical lattices, see pitfalls of pseudo-symmetry in \cite{zwart2008surprises}.
For any small $\ep>0$, the reduced bases $(1,0),(-\ep,2)$ and $(1,0),(-\ep,-2)$ generate the lattices nearly identical to the lattice with the rectangular cell $1\times 2$ (for $\ep=0$), but these bases cannot be made nearly identical by rigid motion, see \citeasnoun[Fig.~4, Corollary~7.9]{kurlin2022mathematics}.
\smallskip

Up to isometry, all real crystals in the CSD are expected to be different, although the recent invariants detected five pairs of suspicious entries, see \citeasnoun[section~7]{widdowson2022average}.
Polymorphs in the CSD, which often have different ref codes, 
can be recognised only by a \emph{continuous} invariant taking similar values for similar crystals. 
\smallskip

Section~\ref{sec:review} reviews the closely related past work.
Section~\ref{sec:root_invariants} reminds the recently developed complete invariants of 2-dimensional lattices. 
Section~\ref{sec:root_maps} maps hundreds of thousands of CSD crystals with full lattice data.
Section~\ref{sec:geo_invariants} explains the geographical metaphor by mapping the invariant values to a sphere, where every lattice (up to rigid motion and uniform scaling) has unique latitude and longitude coordinates.

\section{Overview of key concepts and past work on classifications of lattices}
\label{sec:review}

Crystallography traditionally uses a conventional cell to uniquely represent any periodic crystal, see \citeasnoun[Chapter 2.1]{aroyo2013international}.
In the simpler case of lattices, the cell used is Niggli's reduced cell \cite{niggli1928krystallographische}.
Since this paper studies lattices in $\R^2$, we give the 2-dimensional version obtained from the 3-dimensional definition by choosing a long enough third vector $v_3$ orthogonal to the first two $v_1,v_2$.
For vectors $v_1=(a_1,a_2)$ and $v_2=(b_1,b_2)$ in $\R^2$,  
the determinant of the matrix $\matfour{a_1}{b_1}{a_2}{b_2}$ with the columns $v_1,v_2$ can be defined as $\det(v_1,v_2)=a_1b_2-a_2b_1$.

\begin{dfn}[reduced cell]
\label{dfn:reduced_cell}
For a lattice up to isometry, a basis and its unit cell $U(v_1,v_2)$ are \emph{reduced} (non-acute) if $|v_1|\leq|v_2|$ and $-\frac{1}{2}v_1^2\leq v_1\cdot v_2\leq 0$.
Up to rigid motion, the conditions are weaker: $|v_1|\leq|v_2|$ and $-\frac{1}{2}v_1^2<v_1\cdot v_2\leq\frac{1}{2}v_1^2$, $\det(v_1,v_2)>0$, and the new \emph{special condition} for rigid motion : if $|v_1|=|v_2|$ then $v_1\cdot v_2\geq 0$.
\bs 
\end{dfn}

The conditions for rigid motion did not appear in \citeasnoun[section 9.2.2]{aroyo2013international} because reduced cells were usually considered only up to isometry including reflections.
For example, any rectangular lattice has a unique (up to rigid motion) reduced cell $a\times b$, but two `potentially reduced' bases $v_1=(a,0)$ and $v_2=(0,\pm b)$, which are not related by rigid motion for $0<a<b$. 
Definition~\ref{dfn:reduced_cell} chooses only one of these bases, namely $v_1=(a,0)$ and $v_2=(0,b)$ due to $\det(v_1,v_2)>0$.
\smallskip

Since reduced cells are algorithmically easy to compare up to isometry \cite{kvrivy1976unified}, one can define the discrete metric $d(\La,\La')$ taking the same non-zero value (say, 1) for any non-isometric lattices $\La\not\cong\La'$.
This is the simplest example of a discontinuous metric satisfying all metric axioms on a set of equivalence classes.
\smallskip

Discontinuity of Niggli's basis up to perturbations was practically demonstrated in the seminal work \cite{andrews1980perturbation}.
The introduction of \citeasnoun{edels2021} said that "There is no method for choosing a unique basis for a lattice in a continuous manner. Indeed, continuity contradicts uniqueness as we can continuously deform a basis to a different basis of the same lattice", see a basic example in \citeasnoun[Fig.~3]{kurlin2022mathematics} and a formal proof in \citeasnoun[Theorem~15]{widdowson2022average}.
\smallskip

L.~Andrews and H.~Bernstein have made important advances in \cite{andrews1988lattices,andrews2014geometry,mcgill2014geometry,andrews2019selling} by analysing progressively more complicated boundary cases where cell reductions can be discontinuous.
Since these advances are specialised for $\R^3$, we defer a more detailed review for 3-dimensional lattices to the follow-up paper \cite{bright2021welcome}. 
\smallskip

Another way to represent a lattice $\La\subset\R^n$ is by its Wigner-Seitz cell or Voronoi domain $V(\La)$ consisting of all points $p\in\R^n$ that are closer to the origin $0\in\La$ than to all other points of $\La$.
Though a Voronoi domain $V(\La)$ uniquely determines $\La$ up to rotations, almost any tiny perturbation of a rectangular lattice $\La$ converts the rectangular domain $V(\La)$ into a hexagon.
Hence all combinatorial invariants (numbers of vertices or edges) of $V(\La)$ are discontinuous, similarly in higher dimensions. 
\smallskip

\newcommand{\figuretwo}{
\begin{figure}
\label{fig:Voronoi2D}
\caption{\textbf{Left}: a generic 2D lattice has a hexagonal Voronoi domain with an obtuse superbase $v_1,v_2,v_0=-v_1-v_2$, which is unique up to permutations and central symmetry.
\textbf{Other pictures}: isometric superbases for a rectangular Voronoi domain.}
\includegraphics[width=1.0\textwidth]{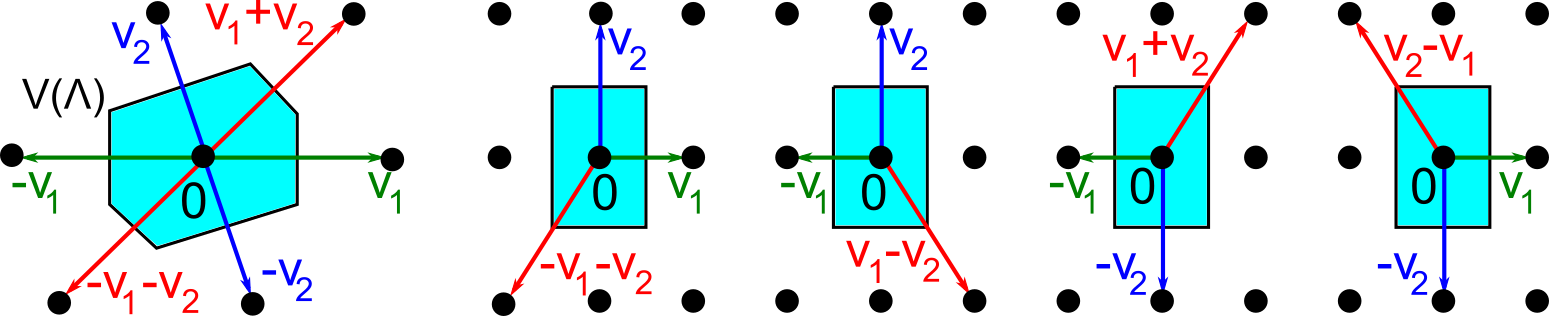}
\end{figure}}
\includefigure{\figuretwo}

However, comparing Voronoi domains as geometric shapes by optimal rotation \cite{mosca2020voronoi} around a common centre led to two continuous metrics on lattices up to rigid motion and similarity.
The minimisation over infinitely many rotations was resolved only by finite sampling, so the exact computation of these metrics is still open.
Similar difficulties remain for general periodic point sets that model all periodic crystals \cite{anosova2021isometry,anosova2021introduction}, which were recently resolved only in dimension 1, see \cite{anosova2022density, kurlin2022computable}.
\smallskip

Another attempt to produce computable metrics was to consider  distance-based invariants \cite{widdowson2022average, widdowson2021pointwise} whose completeness was proved for generic crystals.
These invariants helped establish the Crystal Isometry Principle by experimentally checking that all periodic crystals from the CSD remain non-isometric after forgetting all chemical information.
This principle implies that all periodic crystals can be studied in the common Crystal Isometry Space (CRISP) whose version for 2-dimensional lattices is the Lattice Isometry Space $\LIS(\R^2)$.
\smallskip

The work \cite{conway1992low} came close to mapping lattice spaces without formally stating key challenges, see a full statement in \citeasnoun[Problem~1.1]{kurlin2022mathematics}
\smallskip

\citeasnoun[Proposition 3.10]{kurlin2022mathematics} proves that a reduced basis from Definition~\ref{dfn:reduced_cell} is unique (also up to rigid motion) and all reduced bases are in a 1-1 correspondence with obtuse superbases, which are easier to visualise below, especially for $n\leq 3$.

\begin{dfn}[obtuse superbase, conorms $p_{ij}$]
\label{dfn:conorms}
For any basis $v_1,\dots,v_n$ in $\R^n$, the \emph{superbase} $v_0,v_1,\dots,v_n$ includes the vector $v_0=-\sum\limits_{i=1}^n v_i$.
The \emph{conorms} $p_{ij}=-v_i\cdot v_j$ are the negative scalar products of the vectors above. 
The superbase is called \emph{obtuse} if all conorms $p_{ij}\geq 0$, so all angles between the vectors $v_i,v_j$ are non-acute for distinct indices $i,j\in\{0,1,\dots,n\}$.
The obtuse superbase is \emph{strict} if all $p_{ij}>0$.
\bs
\end{dfn}

Definition~\ref{dfn:conorms} use the conorms $p_{ij}$ from \cite{conway1992low}, which were known as negative Selling parameters \cite{selling1874ueber} and Delone parameters \cite{delone1934mathematical}.
Lagrange \cite{lagrange1773recherches} proved that the isometry class of any lattice $\La\subset\R^2$ with a basis $v_1,v_2$ is determined by the \emph{positive quadratic form} 
$$Q(x,y)=(xv_1+yv_2)^2=q_{11}x^2+2q_{12}xy+q_{22}y^2\geq 0 
\text{ for all }x,y\in\R,$$
where $q_{11}=v_1^2$, $q_{12}=v_1\cdot v_2$, $q_{22}=v_2^2$.
The triple $(v_1^2,v_1\cdot v_2,v_2^2)$ is also called a metric tensor of (a basis of) $\La$. 
Any $Q(x,y)$ has a reduced (non-acute) form with $0<q_{11}\leq q_{22}$ and $-q_{11}\leq 2q_{12}\leq 0$, which is equivalent to reducing a basis up to isometry.
\smallskip

The bases $v_1=(3,0)$, $v_2^{\pm}=(-1,\pm 2)$ generate the mirror images not related by rigid motion, but define the same form $Q=9x^2-6xy+5y^2$ satisfying the reduction conditions above.
So quadratic forms do not distinguish mirror images (enantiomorphs).
Hence the new conditions for the rigid motion were needed in
Definition~\ref{dfn:reduced_cell}.
\smallskip

Motivated by the non-homogeneity of the metric tensor (two squared lengths and scalar product), Delone proposed \cite{delone1937geometry} the homogeneous parameters
$$p_{12}=-v_1\cdot v_2=-q_{12},\quad 
p_{01}=-v_0\cdot v_1=q_{11}+q_{12},\quad  
p_{02}=-v_0\cdot v_2=q_{22}+q_{12},$$
 see all notations in Definition~\ref{dfn:conorms}.
Then any permutation of superbase vectors satisfying $v_0+v_1+v_2=0$ changes \emph{Delone's parameters} $p_{12},p_{01},p_{02}$ by the same permutation of indices.
For example, swapping $v_1,v_2$ is equivalent to swapping $p_{01},p_{02}$.
\smallskip

Delone's reduction \cite{delaunay1973three} proved the key existence result: any lattice in dimensions 2 and 3 has an obtuse superbase with all $p_{ij}\geq 0$.  
Section~\ref{sec:root_invariants} further develops the Delone parameters to show in section~\ref{sec:root_maps} that millions of lattices from real crystals in the CSD nicely distribute in continuous spaces of lattices.

\section{Homogeneous complete invariants of 2D lattices up to four equivalences}
\label{sec:root_invariants}

This section reminds the lattice classifications in Theorem~\ref{thm:classifications2d} based on the recent invariants introduced in Definitions~\ref{dfn:signRI} and~\ref{dfn:PI} from
\citeasnoun[sections~3-4]{kurlin2022mathematics}.

\begin{dfn}[$\sign(\La)$ and root invariants $\RI,\RI^o$]
\label{dfn:signRI}
Let $B=\{v_0,v_1,v_2\}$ be any obtuse superbase of a lattice $\La\subset\R^2$.
If $\La$ is mirror-symmetric (achiral), set $\sign(\La)=0$.
Otherwise $v_0,v_1,v_2$ have different lengths and no right angles, say $|v_1|<|v_2|<|v_0|$.
Then define $\sign(\La)$ as the sign of the determinant $\det(v_1,v_2)$ of the matrix with the columns $v_1,v_2$.
The \emph{root invariant} $\RI(\La)$ is the ordered triple of the root products $r_{ij}=\sqrt{-v_i\cdot v_j}$ for distinct $i,j\in\{0,1,2\}$.
The \emph{oriented root invariant} $\RI^o(\La)$ is $\RI(\La)$ with $\sign(\La)$ as a superscript, which we skip in the case $\sign(\La)=0$ for brevity.
\bs
\end{dfn}

\citeasnoun[Lemma~3.8]{kurlin2022mathematics} proved that $\RI(\La)$ is an isometry invariant of $\La$, independent of an obtuse superbase $B$ because an obtuse superbase of $\La$ is unique up to isometry, also up to rigid motion for non-rectangular lattices.
This uniqueness was missed in \cite{conway1992low} and actually fails in $\R^3$, see \cite{kurlin2022complete}.
\smallskip


\begin{dfn}[projected invariants $\PI,\PI^o$]
\label{dfn:PI}
The root invariants of all lattices $\La\subset\R^2$ live in the triangular cone $\TC$ in Fig.~\ref{fig:TC}.
The triangular projection $\TP:\TC\to\QT$ divides each coordinate by the \emph{size} $\si(\La)=r_{12}+r_{01}+r_{02}$ and projects $\RI(\La)$ to $(\bar r_{12},\bar r_{01},\bar r_{02})$ in the quotient triangle $\QT$ in Fig.~\ref{fig:QT+QS}.
This triangle can be visualised as the isosceles right-angled triangle $\QT=\{x,y\geq 0,\; x+y\leq 1\}\subset\R^2$ parameterised by $x=\bar r_{02}-\bar r_{01}$ and $y=3\bar r_{12}$ 
The resulting pair $\PI(\La)=(x,y)$ is the \emph{projected invariant}.
The oriented invariant $\PI^o(\La)$ is obtained by adding the superscript $\sign(\La)$.
\bs
\end{dfn}

\newcommand{\figurethree}{
\begin{figure}
\label{fig:TC}
\caption{\textbf{Left}: the triangular cone $\TC=\{(r_{12},r_{01},r_{02})\in\R^3 \mid 0\leq r_{12}\leq r_{01}\leq r_{02}\neq 0\}$ represents the space $\RIS$ of all root invariants, see Definition~\ref{dfn:PI}. 
\textbf{Middle}: 
$\TC$ projects to the quotient triangle $\QT$ representing all 2D lattices up to similarity.
\textbf{Right}: $\QT$ is parameterised by $x=\bar r_{02}-\bar r_{01}\in[0,1)$ and $y=3\bar r_{12}\in[0,1]$.
}
\includegraphics[height=40mm]{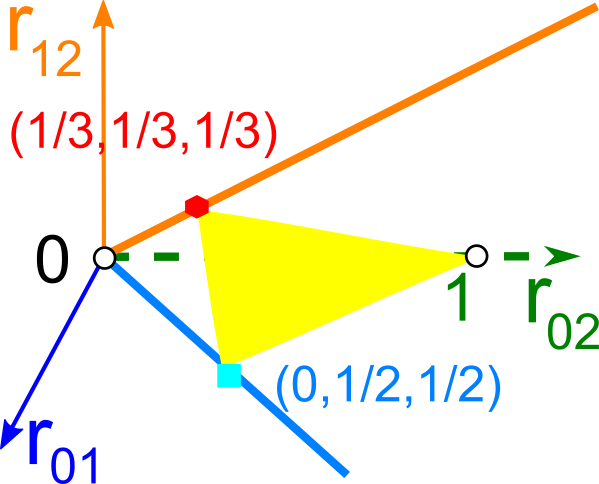}\hspace*{2mm}
\includegraphics[height=40mm]{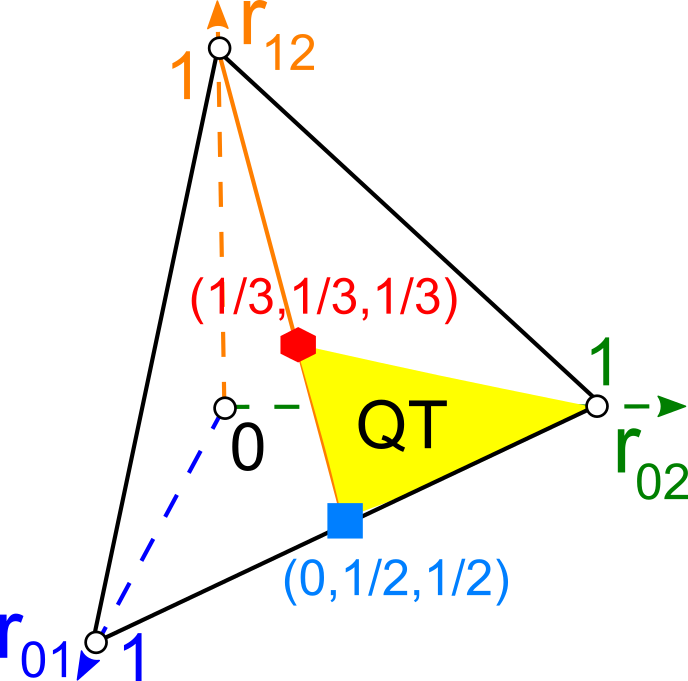}\hspace*{2mm}
\includegraphics[height=40mm]{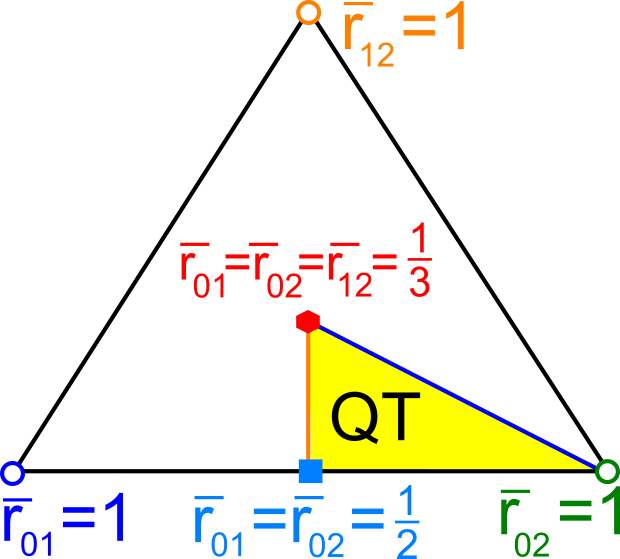}
\end{figure}}
\includefigure{\figurethree}

\newcommand{\figurefour}{
\begin{figure}
\label{fig:QT+QS}
\caption{\textbf{Left}:
all projected invariants $\PI(\La)$ live in the quotient triangle $\QT$ parameterised by $x=\bar r_{02}-\bar r_{01}\in[0,1)$ and $y=3\bar r_{12}\in[0,1]$.
\textbf{Right}: mirror images (enantiomorphs) of any oblique lattice are represented by a pair $(x,y)\lra (1-y,1-x)$ in the quotient square $\QS=\QT^+\cup\QT^-$ symmetric in the diagonal $x+y=1$.}
\includegraphics[width=\textwidth]{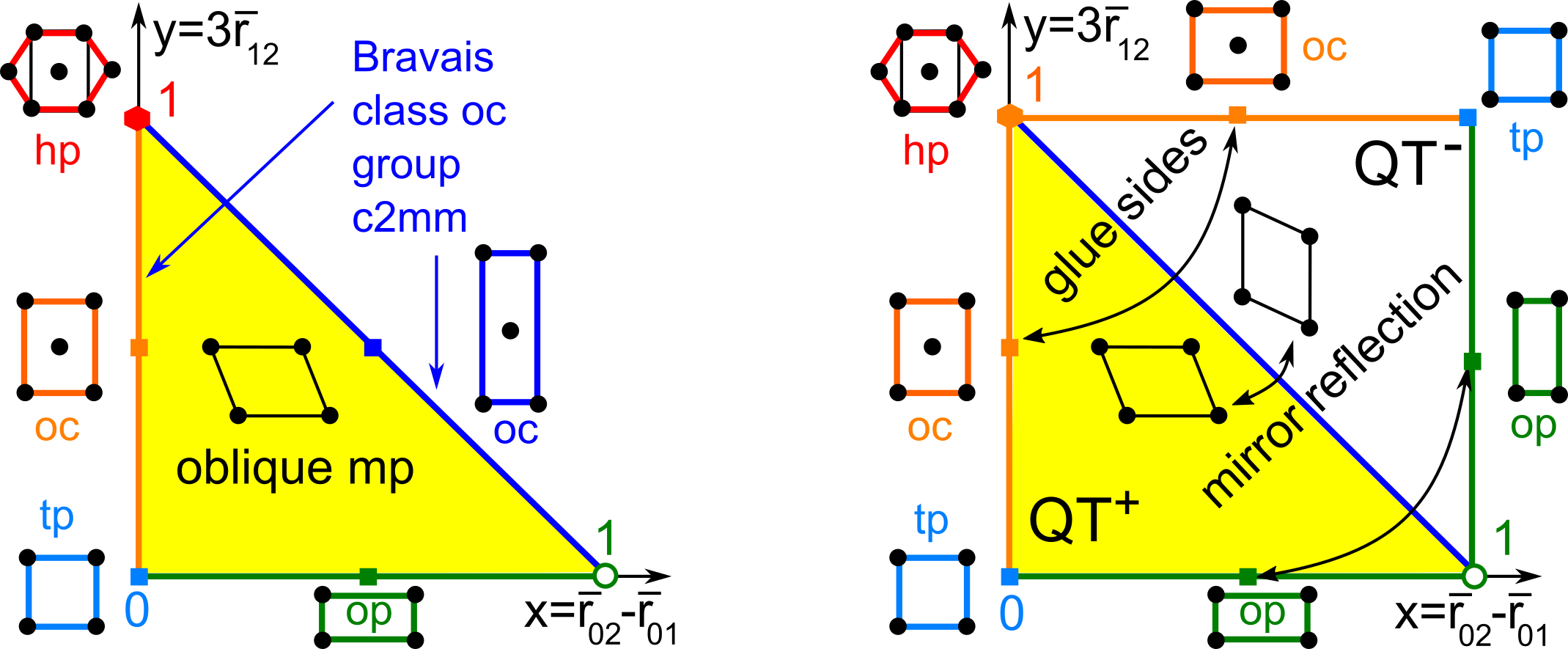}
\end{figure}}
\includefigure{\figurefour}

All oriented projected invariants $\PI^o(\La)$ with $\sign(\La)$ live in a union of two quotient triangles $\QT^+\cup\QT^-$.
These triangles should be glued along the common subspace of mirror-symmetric latices (all non-oblique lattices $\La\subset\R^2$), whose $\PI(\La)$ belong to the boundary of $\QT$.
Fig.~\ref{fig:QT+QS}~(right) glues the hypotenuses of $\QT^{\pm}$ and indicates how to glue the remaining sides.
The gluing produces a topological sphere without a single point due to the excluded vertex $(1,0)$.
We visualise this sphere like the Earth surface with geographic-style coordinates and the boundary of $\QT$ as the equator in section~\ref{sec:geo_invariants}.

\begin{exa}[Bravais classes]
\label{exa:Bravais2d}
\textbf{(tp)}
The square lattice $\La_4\subset\R^2$ with a unit cell $a\times a$ has the root invariant $\RI(\La_4)=(0,a,a)\in\TC$ projected by $\TP$ to $(\bar r_{12},\bar r_{01},\bar r_{02})=(0,\frac{1}{2},\frac{1}{2})$.
By Definition~\ref{dfn:PI} the projected invariant $\PI(\La_4)=(x,y)=(\bar r_{02}-\bar r_{01},3\bar r_{12})=(0,0)\in\QT$, see Fig.~\ref{fig:QT+QS}~(left). 
So the Bravais class (tp) of all square (tetragonal) lattices $\La_4\subset\R^2$ is represented by the bottom-left vertex $(0,0)$ in the quotient triangle $\QT$, identified with the top-right vertex of the quotient square $\QS$ in Fig.~\ref{fig:QT+QS}~(right).
\smallskip

\noindent
\textbf{(hp)}
The hexagonal lattice $\La_6$ with a minimum inter-point distance $a$ has the root invariant $\RI(\La_6)=(\frac{a}{\sqrt{2}},\frac{a}{\sqrt{2}},\frac{a}{\sqrt{2}})$ projected by $\TP$ to $(\frac{1}{3},\frac{1}{3},\frac{1}{3})$.
The projected invariant is $\PI(\La_6)=(x,y)=(0,1)\in\QT$, see Fig.~\ref{fig:QT+QS}~(left).
The Bravais class (hp) of all hexagonal lattices $\La_6\subset\R^2$ is represented by the top-left vertex $(0,1)$ in the quotient triangle $\QT$, identified with the bottom-right vertex of the quotient square $\QS$.
\smallskip

\noindent
\textbf{(op)}
Any rectangular lattice $\La$ with a unit cell $a\times b$ for $0<a<b$ has the obtuse superbase $v_1=(a,0)$, $v_2=(0,b)$, $v_0=(-a,-b)$, see Fig.~\ref{fig:achiral_lattices}~(left).
Then $\RI(\La)=(0,a,b)$ and $\PI(\La)=(\frac{b-a}{b+a},0)$ belongs to the horizontal side of $\QT$, which represents the Bravais class (op). 
We approach the excluded vertex $(1,0)$ as $b\to+\infty$.
\smallskip

\noindent
\textbf{(oc)}
Any centred rectangular lattice $\La$ with a conventional unit cell $2a\times 2b$ for $0<a<b$ has the obtuse superbase $v_1=(2a,0)$, $v_2=(-a,b)$, $v_0=(-a,-b)$, see Fig.~\ref{fig:achiral_lattices}.
Then $r_{01}=a\sqrt{2}=r_{02}$ and $r_{12}=\sqrt{b^2-a^2}$.
If $b\leq a\sqrt{3}$, then $\RI(\La)=(\sqrt{b^2-a^2},a\sqrt{2},a\sqrt{2})$ and $\PI(\La)=(0,\frac{3\sqrt{b^2-a^2}}{2a\sqrt{2}+\sqrt{b^2-a^2}})$ belongs to the vertical side of $\QT$.
For $b=a$, the lower vertex $(x,y)=(0,0)$ of $\QT$ represents all square lattices with $r_{12}=0$.
For $b=a\sqrt{3}$, the upper vertex $(x,y)=(0,1)$ of $\QT$ represents all hexagonal lattices with $r_{12}=r_{01}=r_{02}$.
If $b>a\sqrt{3}$, then $\RI(\La)=(a\sqrt{2},a\sqrt{2},\sqrt{b^2-a^2})$ and $\PI(\La)=(\frac{3a\sqrt{2}}{2a\sqrt{2}+\sqrt{b^2-a^2}},\frac{\sqrt{b^2-a^2}-a\sqrt{2}}{2a\sqrt{2}+\sqrt{b^2-a^2}})$ belongs to the hypotenuse $x+y=1$ of $\QT$.
The open vertical edge and open hypotenuse of $\QT$ represent the Bravais class $oc$ of all centred rectangular lattices.
The excluded vertex $(1,0)$ represents the limit $b\to+\infty$.
\bs
\end{exa}

\newcommand{\figurefive}{
\begin{figure}
\label{fig:achiral_lattices}
\caption{
\textbf{Left}: any rectangular lattice $\La$ with a unit cell $a\times b$ has the obtuse superbase $B$ with $v_1=(a,0)$, $v_2=(0,b)$, $v_0=(-a,-b)$, see Example~\ref{exa:Bravais2d}(op).
Other lattices $\La$ have a rectangular cell $2a\times 2b$ and an obtuse superbase $B$ with $v_1=(2a,0)$, $v_2=(-a,b)$, $v_0=(-a,-b)$.
\textbf{Middle}: $\RI(\La)=(\sqrt{b^2-a^2}, a\sqrt{2},a\sqrt{2})$, $a\leq b\leq a\sqrt{3}$.
\textbf{Right}: $\RI(\La)=(a\sqrt{2},a\sqrt{2},\sqrt{b^2-a^2})$, $a\sqrt{3}\leq b$, see Example~\ref{exa:Bravais2d}(oc).}
\includegraphics[width=1.0\textwidth]{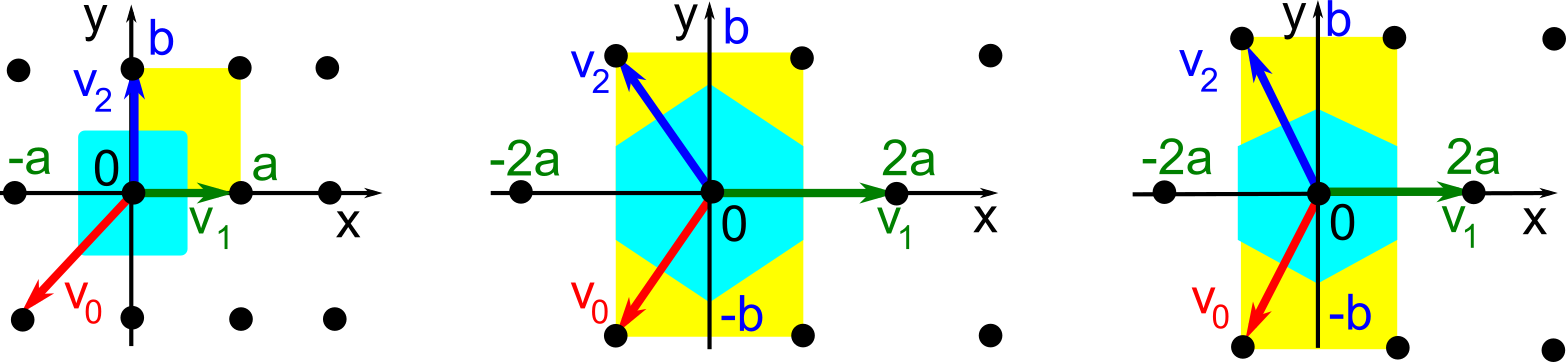}
\end{figure}
}
\includefigure{\figurefive}

Theorem~\ref{thm:classifications2d} summarises the complete classifications of 2-dimensional lattices from
\citeasnoun[Theorem~4.2 and Corollary~4.6]{kurlin2022mathematics} 
up to four equivalence relations.

\begin{thm}[lattice classifications]
\label{thm:classifications2d}
Let $\La,\La'\subset\R^2$ be any lattices. 
\smallskip

\noindent
\textbf{(a)}
$\La,\La'$ are isometric if and only if $\RI(\La)=\RI(\La')$;
\smallskip

\noindent
\textbf{(b)}
$\La,\La'$ are similar if and only if $\PI(\La)=\PI(\La')$;
\smallskip

\noindent
\textbf{(c)}
$\La,\La'$ are related by rigid motion if and only if $\RI^o(\La)=\RI^o(\La')$;
\smallskip

\noindent
\textbf{(d)}
$\La,\La'$ are similar with the same orientation if and only if $\PI^o(\La)=\PI^o(\La')$.
\bs
\end{thm}

\section{Mapping millions of 2-dimensional lattices extracted from CSD crystals}
\label{sec:root_maps}

For any periodic crystal from the Cambridge Structural Database (CSD), which has full geometric data of its lattice $\La\subset\R^3$, we extract three 2-dimensional lattices generated by three pairs $\{v_2,v_3\}$, $\{v_1,v_3\}$, $\{v_1,v_2\}$ of given basis vectors of $\La$.
\smallskip

Fig.~\ref{fig:CSD_QS} shows all resulting 2.6 million lattices in the quotient square $\QS$.
Only about 55\% of all lattices have non-trivial symmetries of Bravais classes oc, op, hp, tp.
The remaining 45\% of lattices are oblique and nicely fill $\QS$ apart from the lower right corner where primitive unit cells are rather elongated.
Hence we conclude that the Lattice Isometry Space $\LIS(\R^2)$ is continuously populated by real CSD crystals.

\newcommand{\figuresix}{
\begin{figure}
\label{fig:CSD_QS}
\caption{
Density maps in $\QS$ of all 2D lattices extracted from CSD crystals.
The colour of each pixels indicates (on the logarithm scale) the number of lattices whose projected invariant $\PI(\La)=(x,y)=(\bar r_{02}-\bar r_{01},3\bar r_{12})$ belongs to this pixel.
The darkest pixels represent rectangular lattices on the bottom and right edges of $\QS$. 
}
\includegraphics[width=\textwidth]{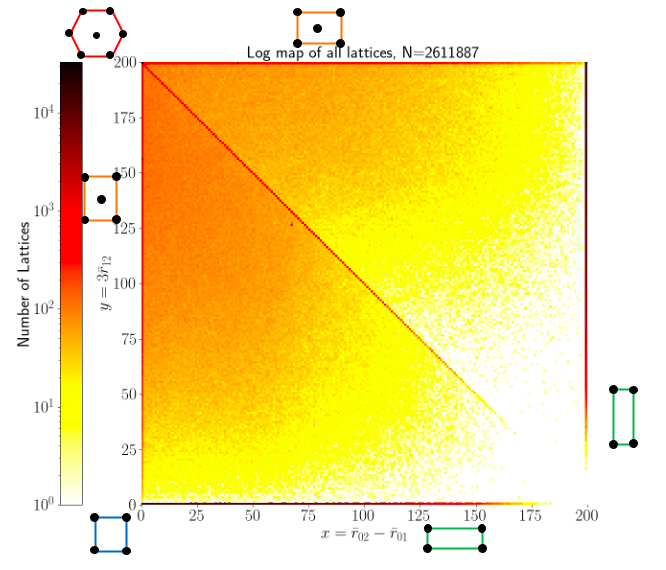}
\end{figure}}
\includefigure{\figuresix}

\newcommand{\figureseven}{
\begin{figure}
\label{fig:CSD_QS_oblique}
\caption{
Normal density map all 2D oblique lattices from CSD crystals in $\QS$.
After removing mirror-symmetric lattices on the boundary and diagonal of $\QS$, we can better see the tendency towards hexagonal lattices at the top left corner $(0,1)\in\QS$.}
\includegraphics[width=\textwidth]{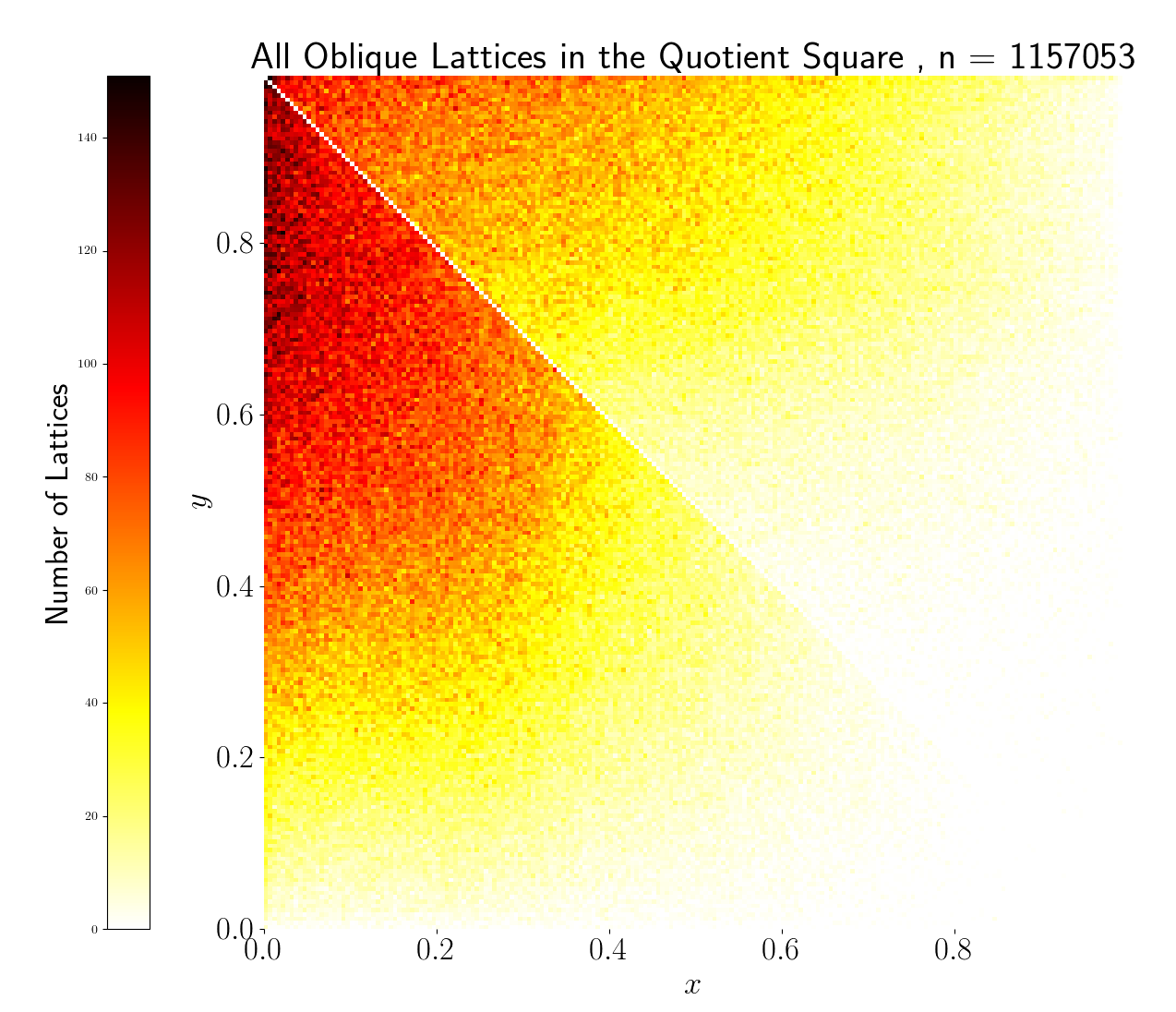}
\end{figure}}
\includefigure{\figureseven}

After removing all non-oblique lattices represented by root invariants along the sides and the diagonal of $\QS$, the map in Fig.~\ref{fig:CSD_QS_oblique} shows the preference for positive lattices which can be explained by a standard order of basis vectors for monoclinic lattices.
 
\newcommand{\figureeight}{
\begin{figure}
\label{fig:CSD_rectangular}
\caption{
Density maps of parameters $(a,b)$ in $\AA$. 
\textbf{Left}: rectangular lattices with primitive unit cells $a\times b$ in $N=1094109$ crystals in the CSD.
\textbf{Right}: centred-rectangular lattices with conventional cells $2a\times 2b$ in $N=147451$ crystals in the CSD.}
\includegraphics[width=0.49\textwidth]{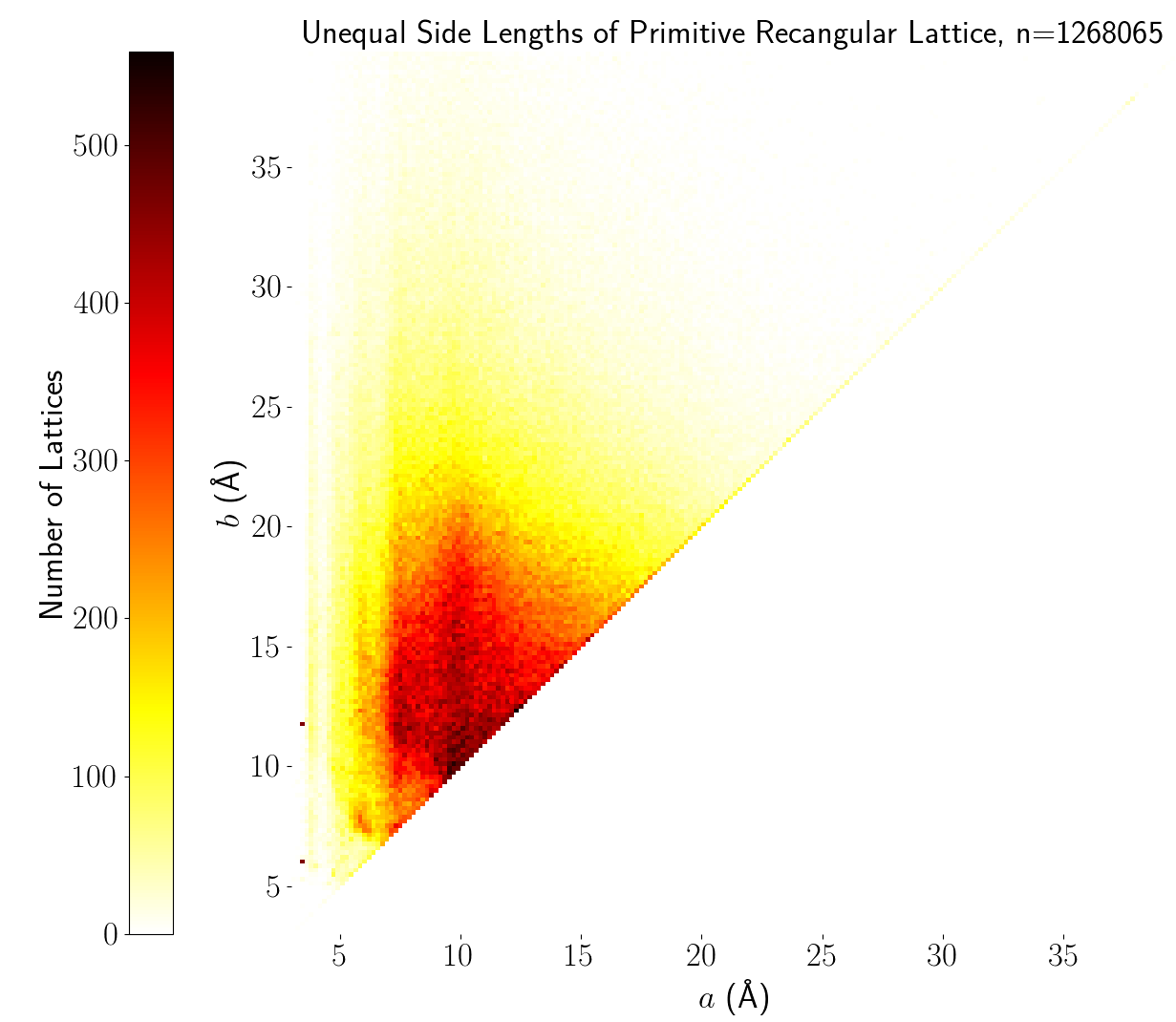}
\includegraphics[width=0.49\textwidth]{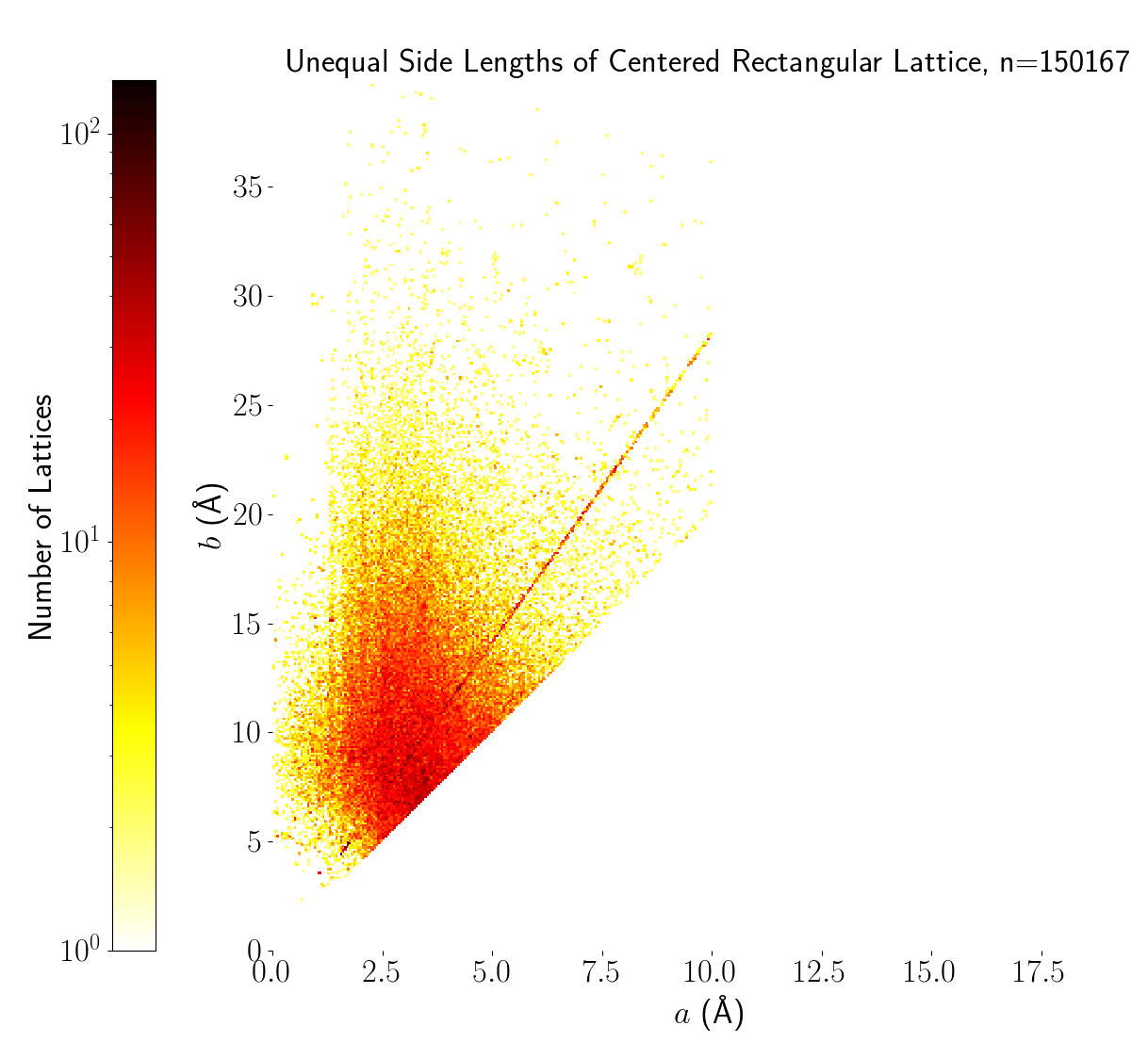}
\end{figure}}
\includefigure{\figureeight}

The density map of rectangular lattices in Fig.~\ref{fig:CSD_rectangular}~(left) has 
 two high density (black) pixels at $a \approx 3.5$ Angstroms arising from  
 $386$ near-identical primitive monoclinic crystals of $\alpha$-oxalic acid dihydrate. 
 This molecule has been used as a benchmark for the calculation of electron densities since its crystallographic properties were thoroughly documented in~\cite{stevecop1980oxalic}.
As a result, hundreds of publications have since generated and deposited further refinements of its structural determination. 
\smallskip

In the density map of centered rectangular lattices in Fig.~\ref{fig:CSD_rectangular}~(right), the most prominent features are the high-density area in the region where the shortest side length is between $2.5$ and $5$ $\AA$.
We also see a visible line $b=\sqrt{2}a$ of high-density pixels. 
This line represents 2D lattices in body-centered cubic lattices, where the ratio of side lengths is $\sqrt{2}$. 
Another high-density pixel in this line
represents $130$ structures of a standard test molecule (hexamethylenetetramine), which was frequently used in the investigation of lattice vibrations~\cite{becka1963hex}. 

\newcommand{\figurenine}{
\begin{figure}
\label{fig:CSD_hex_square}
\caption{
The histograms of minimum inter-point distances $a$ in Angstroms.
\textbf{Left}: all hexagonal 2D lattices in CSD crystals.
\textbf{Right}: all square 2D lattices in CSD crystals.
}
\includegraphics[width=0.49\textwidth]{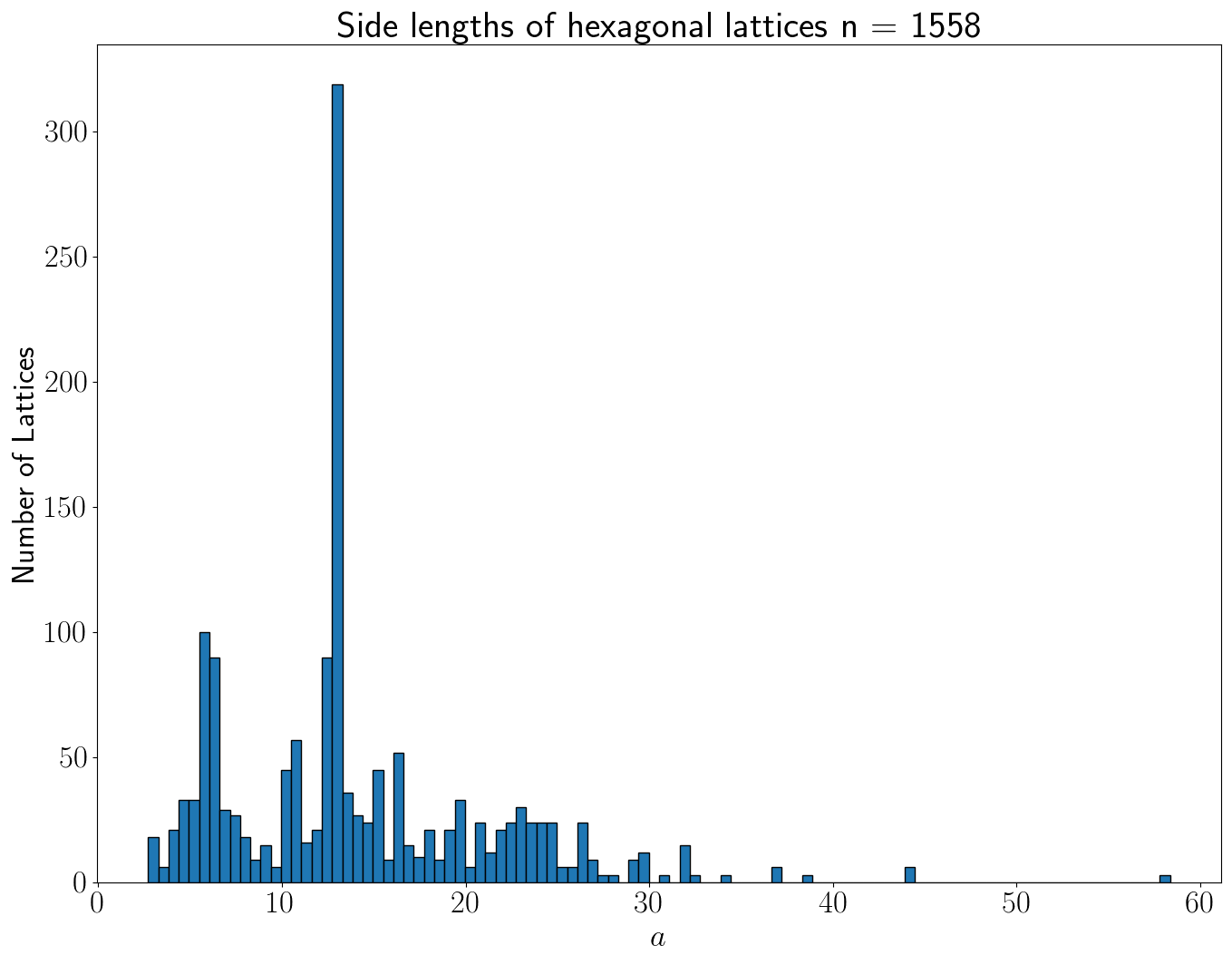}
\includegraphics[width=0.49\textwidth]{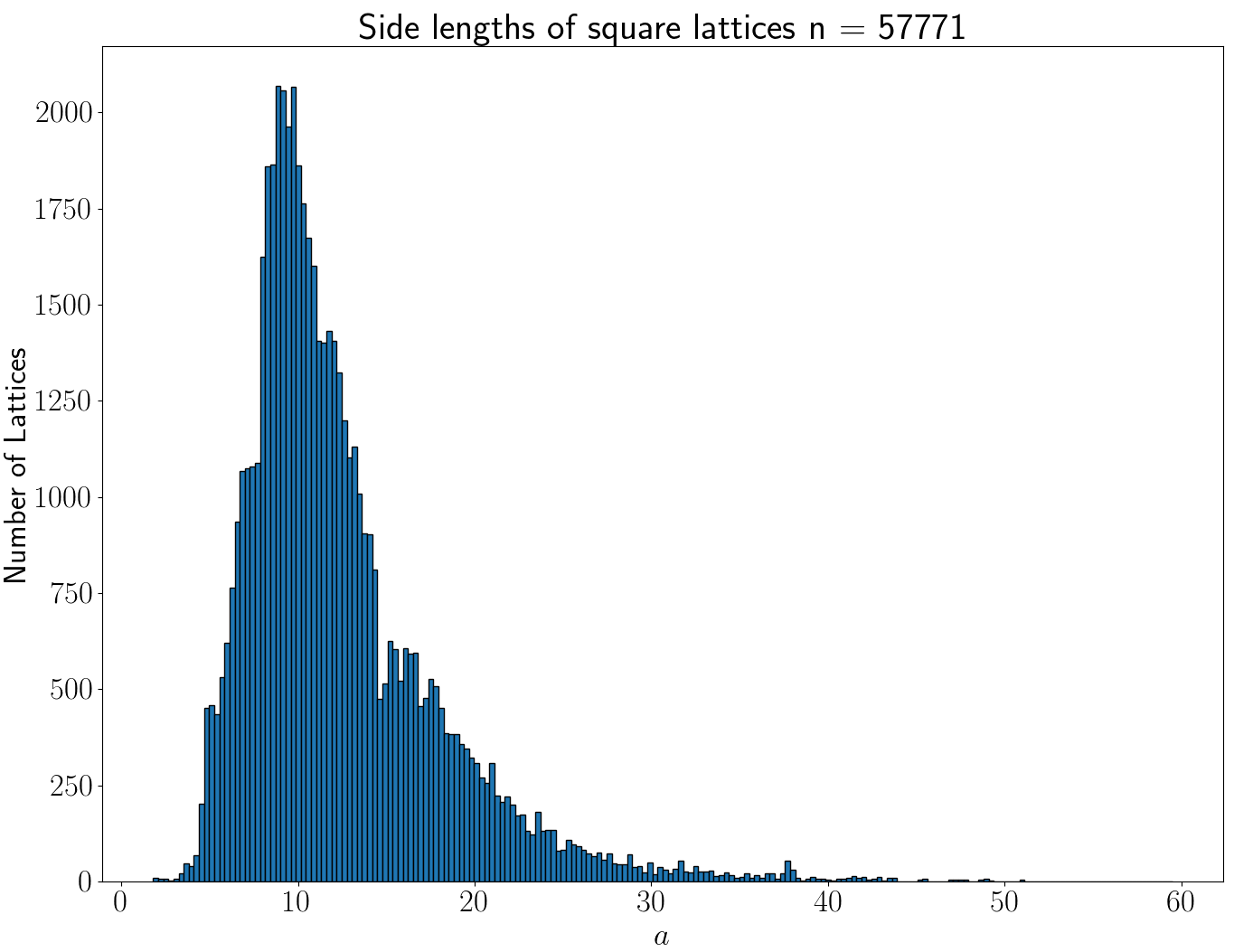}
\end{figure}}
\includefigure{\figurenine}

\section{Other complete invariants and a spherical map of 2-dimensional lattices }
\label{sec:geo_invariants}

The Lattice Isometry/Similarity Spaces can be parameterised in many different ways.
The root invariant $\RI(\La)$ of a lattice $\La\subset\R^2$ has the advantage of homogeneity in the sense that any permutation $\si$ of (indices of) superbase vectors $v_0,v_1,v_2$ permutes the three root products accordingly: $r_{ij}\mapsto r_{\si(i)\si(j)}$.
The metric tensor $\MT=(v_1^2,v_1\cdot v_2,v_2^2)$ including the coefficients of the form $Q_\La(x,y)=q_{11}x^2+2q_{12}xy+q_{22}y^2$ representing $\La$ is not homogeneous in the above sense.
Taking square roots gives the \emph{quadratic invariant} $\QI(\La)=(\tau_{11},\tau_{12},\tau_{22})=(\sqrt{q_{11}},\sqrt{-q_{12}},\sqrt{q_{22}})$ in the units of basis coordinates.
\smallskip

The quadratic invariant $\QI(\La)$ is complete up to isometry by Theorem~\ref{thm:classifications2d}(a) and can be extended to the case of rigid motion by using $\sign(\La)$ from Definition~\ref{dfn:signRI},
\smallskip

In the isosceles triangle $\QT$, continuous metrics and chiral distances have simpl formulae in \citeasnoun[sections 5-6]{kurlin2022mathematics} for the coordinates $x=\bar r_{02}-\bar r_{01}$, $y=3\bar r_{12}$ but can be now re-written for any coordinates on $\LIS(\R^2)$, see the earlier non-isosceles triangles in \citeasnoun[Fig.~1.2 on p.~82]{engel2004lattice} and \citeasnoun[Fig.~6.2]{zhilinskii2016introduction}.
\smallskip

Since the quotient square $\QS=\QT^+\cup\QT^-$ with identified sides is a topological sphere without a single point, it is natural to visualise $\QS$ as the round surface of Earth with $\QT^\pm$ as the north/south hemispheres separated by the equator along their common boundary of $\QT$ represented by $\PI(\La)$ of all mirror-symmetric lattices $\La$.
\smallskip

We can choose any internal point of the quotient triangle $\QT$ as the north pole.
The most natural choice is the incentre $P^+$ (pole), the centre of the circle inscribed into $\QT^+$ because the rays from $P^+$ to the vertices of $\QT^+$ nicely bisect the angles $90^\circ, 45^\circ,45^\circ$. 
The incentre of $\QT^+$ has the coordinates $(x,x)$, where $x=1-\frac{1}{\sqrt{2}}=\frac{1}{2+\sqrt{2}}$.
The lattice $\La_2^+$ with the projected invariant $\PI(\La_2^+)=(x,x)$ has the basis $v_1\approx(1.9,0)$, $v_2\approx(-0.18,3.63)$ inversely designed in \citeasnoun[Example~4.10~($\La_2$)]{kurlin2022mathematics}.

\newcommand{\figureten}{
\begin{figure}
\label{fig:QT+HS}
\caption{
\textbf{Left}: in $\QT^+$, the Greenwich line goes from the `empty' point (1,0) through incentre $P^+$ to the point $G=(0,\sqrt{2}-1)$.
\textbf{Middle}: the hemisphere $\HS^+$ has the north pole at $P^+$, the equator $\bd\QT^+$ of mirror-symmetric lattices.
\textbf{Right}: the longitude $\mu\in(-180^\circ,+180^\circ]$ anticlockwise measures angles from the Greenwich line, the latitude $\ph\in[-90^\circ,+90^\circ]$ measures angles from the equator to the north.}
\includegraphics[height=42mm]{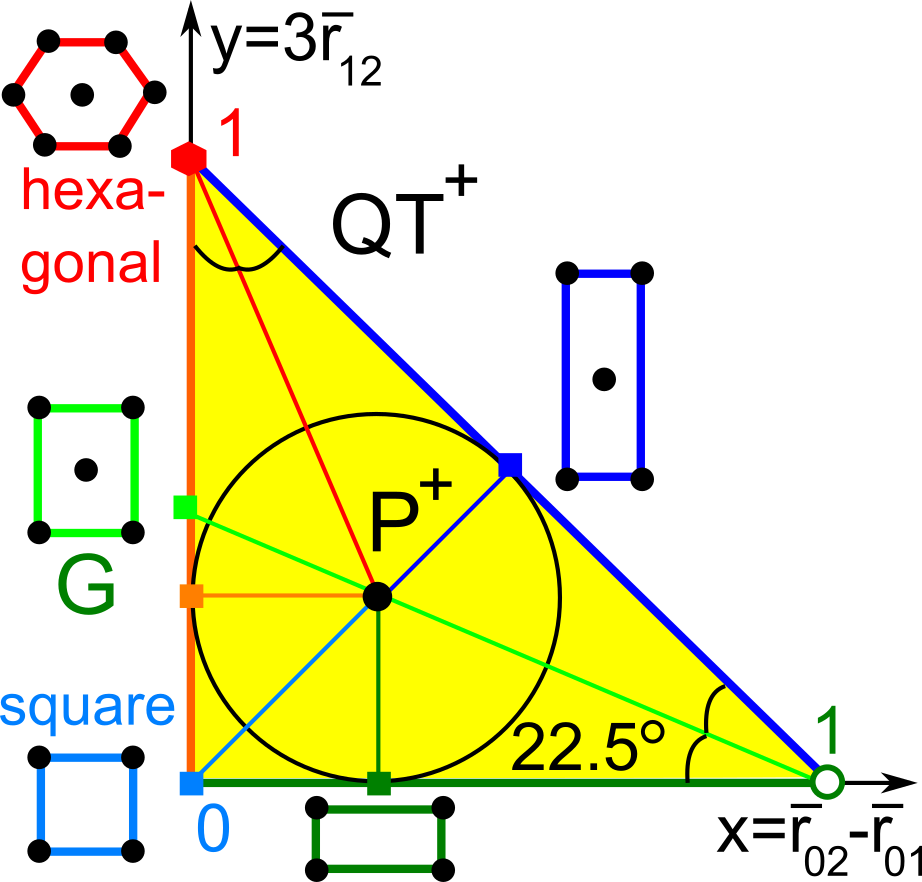}
\includegraphics[height=42mm]{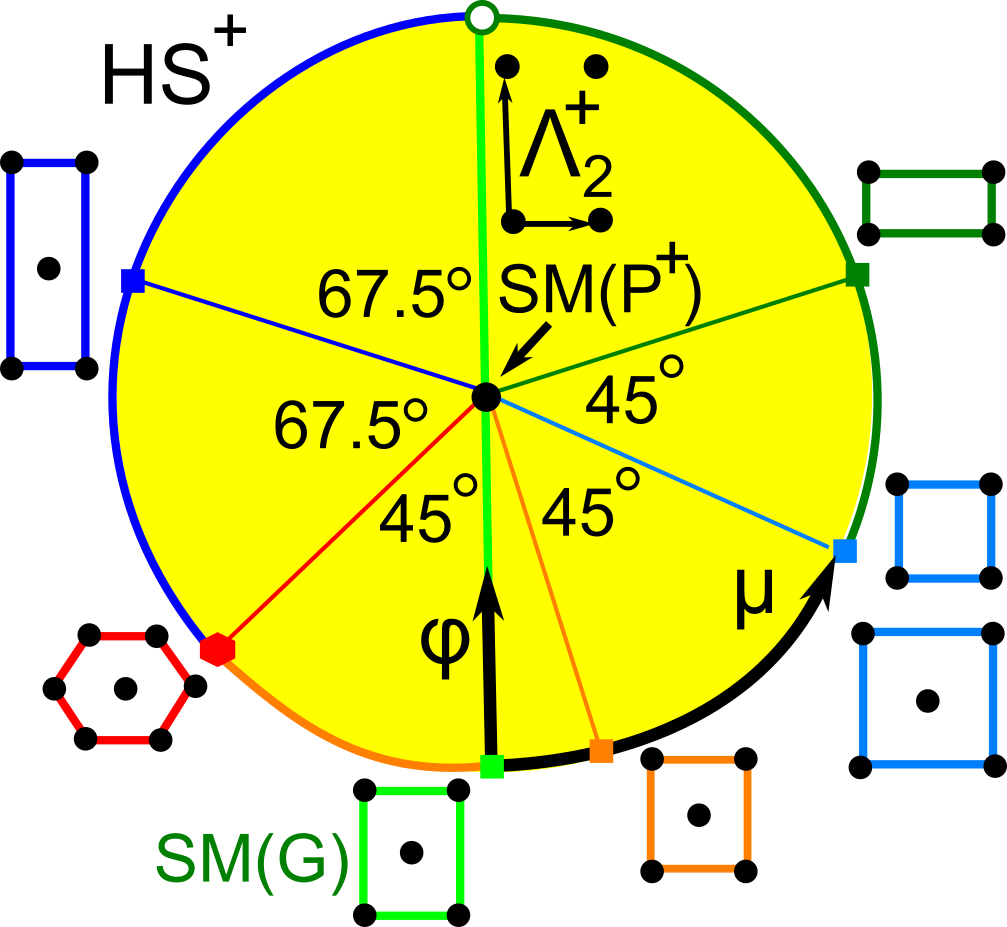}
\hspace*{1mm}
\includegraphics[height=42mm]{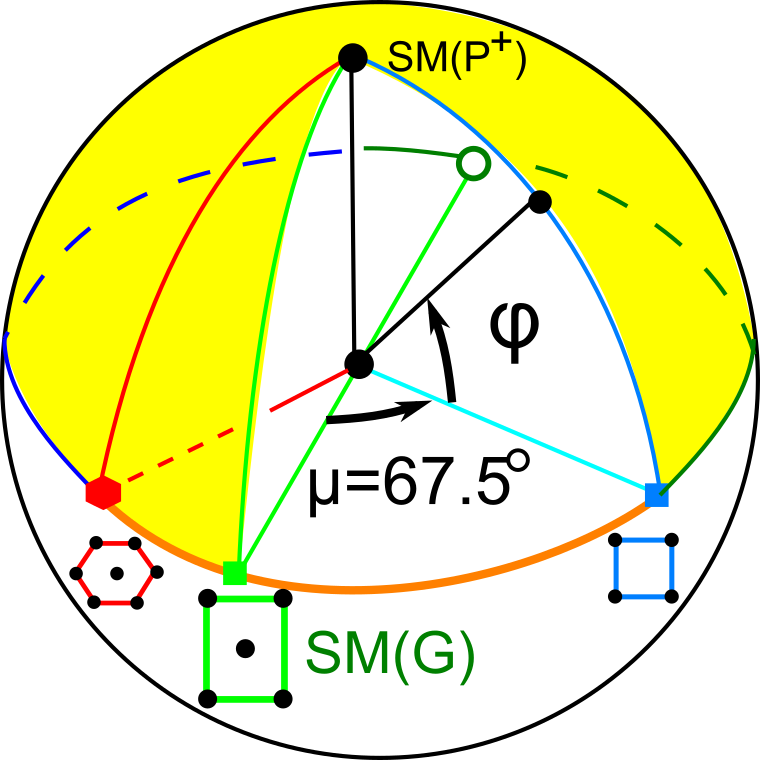}
\end{figure}}
\includefigure{\figureten}

\begin{dfn}[spherical map $\SM:\QS\to S^2$] 
\label{dfn:SM}
\textbf{(a)}
The \emph{spherical map} $\SM$ sends the incentre $P^+$ of $\QT$ to the north pole of the hemisphere $\HS^+$ and the boundary $\bd\QT$ to the equator of $\HS^+$, see Fig.~\ref{fig:QT+HS}~(middle).
Linearly map the line segment between $P^+$ and
any point $(x,y)$ in the boundary $\bd\QT$ to the shortest arc connecting the north pole $\SM(P^+)$ to $\SM(x,y)$ in the equator of $\HS^+$.
Extend the \emph{spherical map} to $\SM:\QS\to S^2$ by sending any pair of invariants $\PI^o(\La^{\pm})$ with $\sign(\La^{\pm})=\pm 1$ to the northern/southern hemispheres of the 2-dimensional sphere $S^2$, respectively.
\smallskip

\noindent
\textbf{(b)}
For any lattice $\La\subset\R^2$, the \emph{latitude} $\ph(\La)\in[-90^\circ,+90^\circ]$ is the angle from the equatorial plane $\EP$ of $S^2$ to the radius-vector to the point $\SM(\PI^o(\La))\in S^2$ in the upwards direction.
Let $v(\La)$ be the orthogonal projection of this radius-vector to $\EP$.
Let the \emph{Greenwich meridian} be the spherical arc going through $\SM(G)$ in the equator $E$, where the \emph{Greenwich} point $G=(0,\sqrt{2}-1)\in\bd\QT$ is in the line through $P^+$ and $(1,0)$.
The longitude $\mu(\La)\in(-180^\circ,180^\circ]$ is the anticlockwise angle from the \emph{Greenwich plane} through the Greenwich meridian to the vector $v(\La)$ above.
\bs
\end{dfn}

For lattices with $\PI(\La)$ in the straight-line segment between the excluded vertex $(1,0)$ and the incentre $P^+$, we choose the longitude $\mu=+180^\circ$ rather than $-180^\circ$.
Proposition~\ref{prop:SM} computes 
$\mu(\La),\ph(\La)$ via $\PI(\La)=(x,y)$
and is proved in Appendix~\ref{sec:metrics}.

\begin{prop}[formulae for $\SM$]
\label{prop:SM}
For any lattice $\La\subset\R^2$ with 
$\PI(\La)=(x,y)\in\QT$,
 if $x\neq t=1-\dfrac{1}{\sqrt{2}}$, then set $\psi=\arctan\dfrac{y-t}{x-t}$, otherwise $\psi=\sign(y-t)90^\circ$.

\noindent
(\ref{prop:SM}a)
The longitude of the lattice $\La$ is 
$\mu(\La)=\left\{\begin{array}{l} 
\psi+22.5^\circ \text{ if } x<t, \\
\psi-157.5^\circ \text{ if } x\geq t, \psi\geq-22.5^\circ, \\
\psi+202.5^\circ \text{ if } x\geq t, \psi\leq-22.5^\circ.
\end{array} \right.$

\noindent
(\ref{prop:SM}b)
The latitude is 
$\ph(\La)=\sign(\La)\cdot\left\{\begin{array}{l} 
\frac{x\sqrt{2}}{\sqrt{2}-1}90^\circ \text{ if } \mu(\La)\in[-45^\circ,+67.5^\circ], \\
\frac{y\sqrt{2}}{\sqrt{2}-1}90^\circ \text{ if }\mu(\La)\in[+67.5^\circ,+180^\circ], \\
\frac{1-x-y}{\sqrt{2}-1}90^\circ \text{ if } \mu(\La)\in[-180^\circ,-45^\circ].
\end{array} \right.$

\noindent
The incentres $P^{\pm}\in\QT^{\pm}$ have $\psi=0$ and $\mu=\pm 90^{\circ}$, respectively, $\ph$ is undefined.
\bs
\end{prop}

\begin{exa}[prominent lattices]
\label{exa:SM}
Any mirror-symmetric lattice $\La\subset\R^2$ has $\sign(\La)=0$, hence belongs to the equator $E$ of $S^2$ and has latitude $\ph(\La)=0$ by (\ref{prop:SM}b).
Any square lattice $\La_4$ with $\PI(\La_4)=(0,0)$ has $\mu(\La_4)=\arctan 1+22.5^\circ=67.5^\circ$ by (\ref{prop:SM}a).
Any hexagonal lattice $\La_6$ with $\PI(\La_4)=(0,1)$ has $\mu(\La_4)=\arctan\frac{1}{1-\sqrt{2}}+22.5^\circ=-45^\circ$.
Any rectangular lattice $\La$ with $\PI(\La)=(1-\frac{1}{\sqrt{2}},0)$ has $\mu(\La)=-90^\circ+202.5^\circ =112.5^\circ$.
Any centered rectangular lattice $\La$ with $\PI(\La)=(\frac{1}{2},\frac{1}{2})$ at the mid-point of the diagonal of $\QT$ has $\mu(\La)=\arctan 1-157.5^\circ =-112.5^\circ$.
Any \emph{Greenwich} lattice $\La_G$ with $\PI(\La_G)=G=(0,\sqrt{2}-1)$ has $\mu(\La_G)=\arctan(1-\sqrt{2})+22.5^\circ=0$.
\bs
\end{exa}

Figures~\ref{fig:snorth},\ref{fig:ssouth},\ref{fig:swest},\ref{fig:seast} show the density maps of 2D lattices from Figure~\ref{fig:CSD_QS} on the northern, southern, western, eastern hemispheres for the spherical map $\SM:\QS\to S^2$.

\newcommand{\figureeleven}{
\begin{figure}
\label{fig:snorth}
\caption{
The density map of 2D lattices from CSD crystals on the northern hemisphere. 
The circumference (equator) is parameterised by the longitude $\mu\in (-180^\circ,180^\circ]$.
The radial distance is the latitude $\phi\in [0^\circ,90^\circ]$.
\textbf{Left}: all $N=2191887$ lattices with $\sign(\La)\geq 0$, $\phi \geq 0$. 
\textbf{Right}: all $N=741105$ oblique lattices with $\sign(\La)> 0$, $\phi > 0$.}
\includegraphics[width=0.49\textwidth]{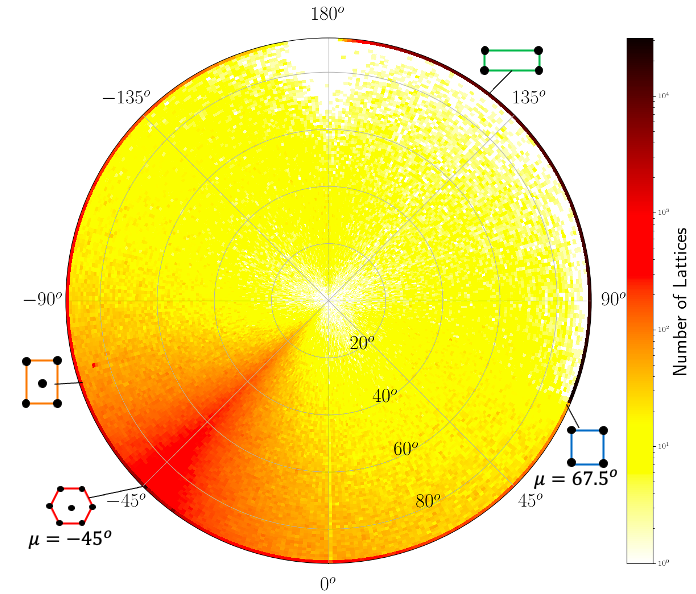}
\includegraphics[width=0.49\textwidth]{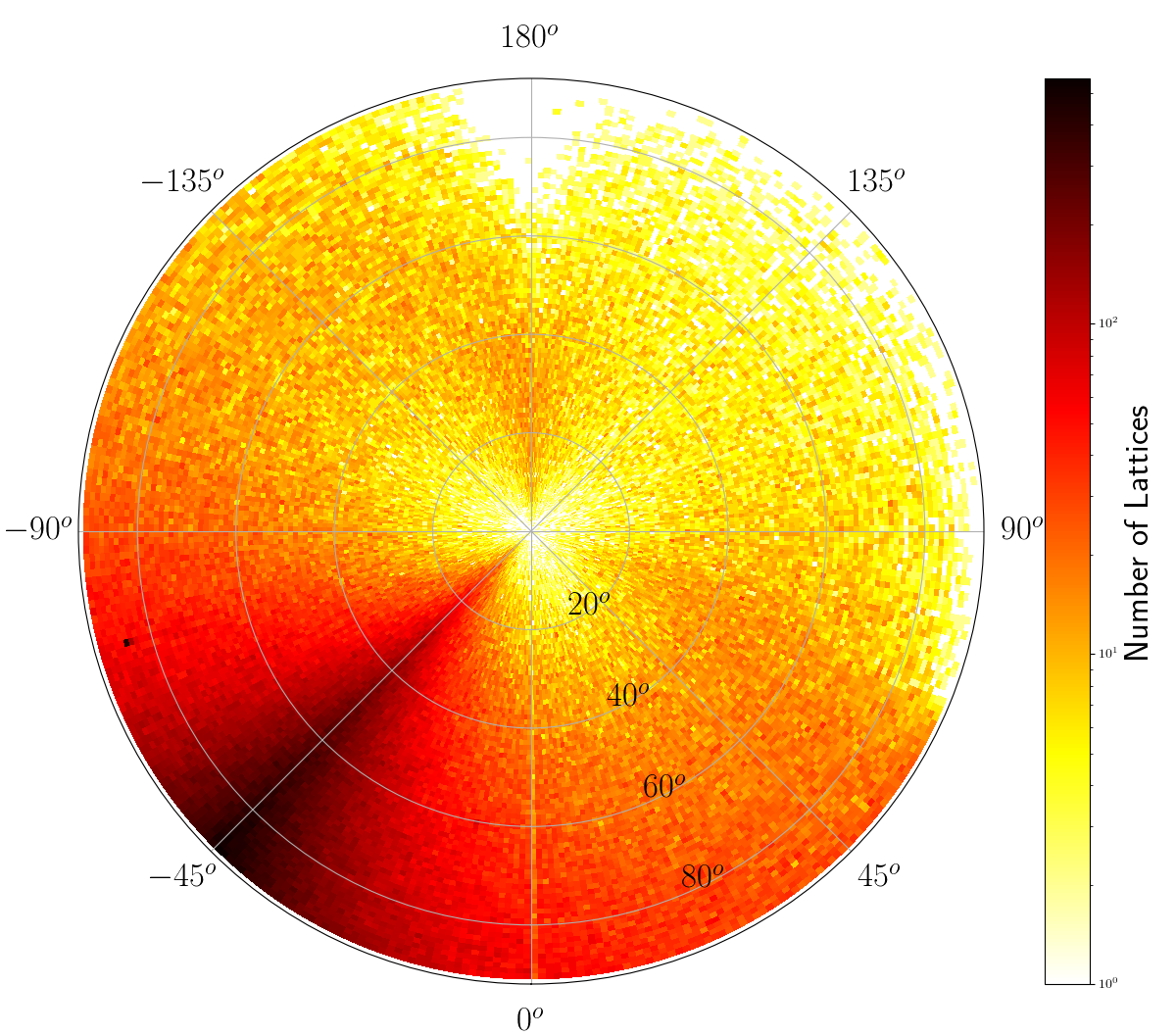}
\end{figure}}
\includefigure{\figureeleven}

\newcommand{\figuretwelve}{
\begin{figure}
\label{fig:ssouth}
\caption{
The density map of 2D lattices from CSD crystals on the northern hemisphere. 
The circumference (equator) is parameterised by the longitude $\mu\in (-180^\circ,180^\circ]$.
The radial distance is the latitude $\phi\in [0^\circ,90^\circ]$.
\textbf{Left}: all $N=1854209$ lattices with $\sign(\La)\leq 0$, $\phi \leq 0$.
\textbf{Right}: all $N=406930$ oblique lattices with $\sign(\La)<0$, $\phi<0$.}
\includegraphics[width=0.49\textwidth]{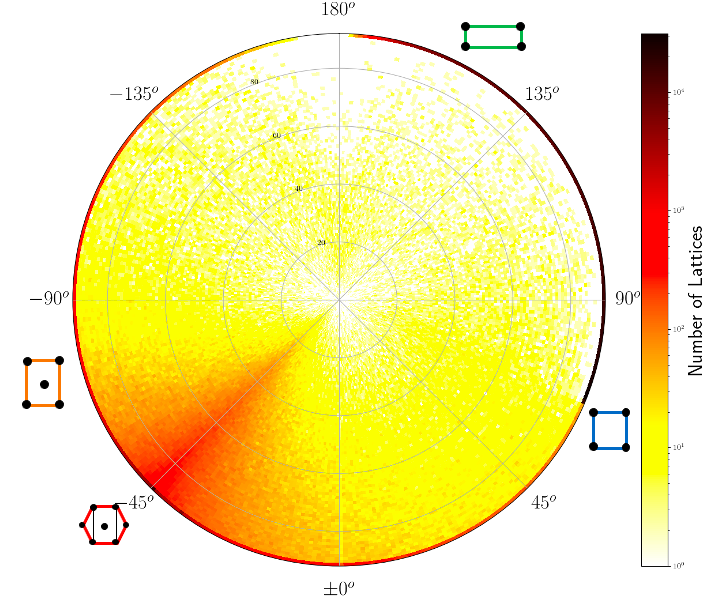}
\includegraphics[width=0.49\textwidth]{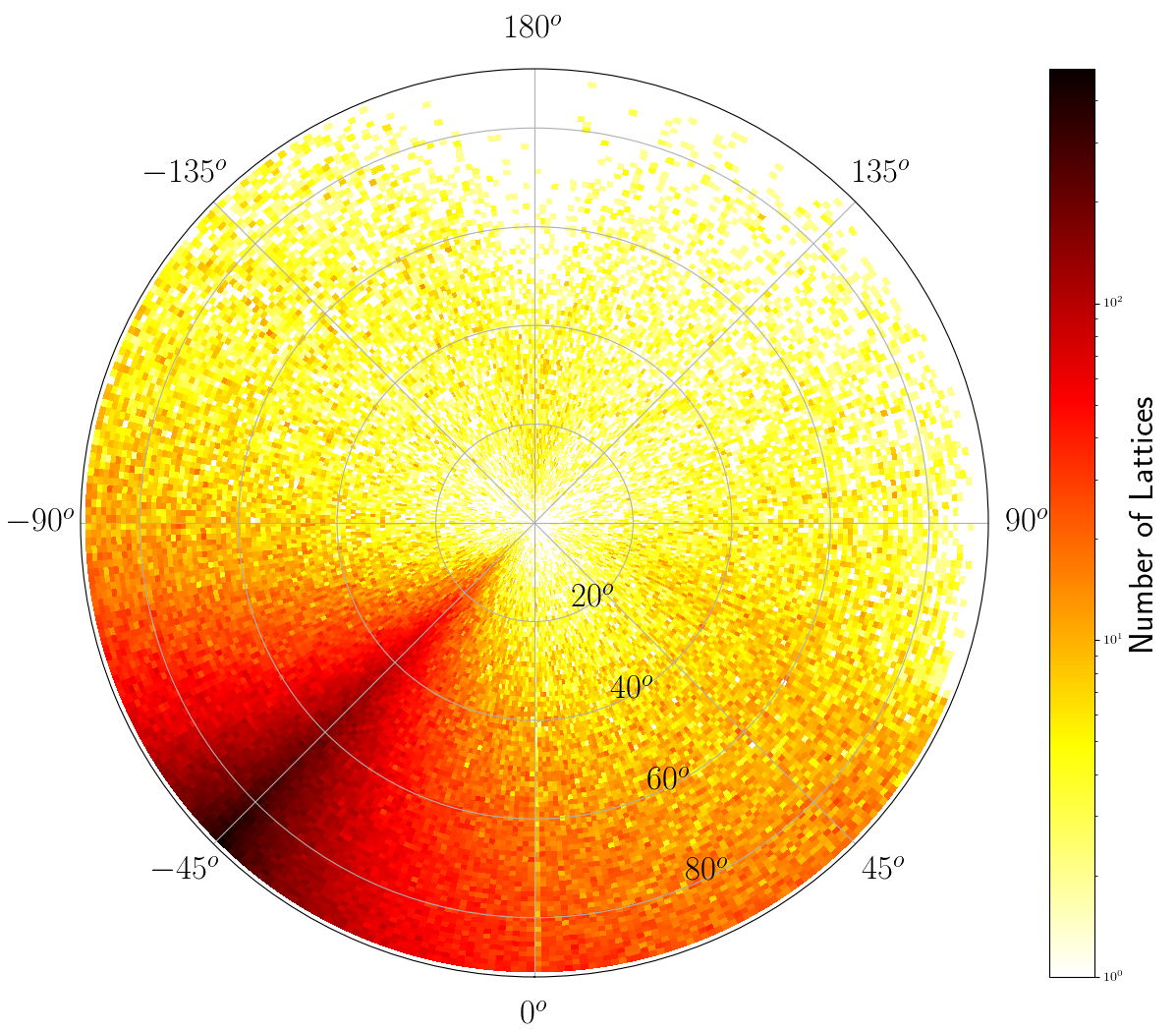}
\end{figure}}
\includefigure{\figuretwelve}

The north pole represents the incentre $P^{+}$ whose pixel contains $230$ lattices in Figure~\ref{fig:CSD_QS_oblique} but appears sparsely populated in Fig.~\ref{fig:snorth} because this incentre pixel is split into many more $1\times 1$ degree curved `pixels' of a much lower density. 
The high density near the point representing hexagonal lattices is visible in Figures~\ref{fig:snorth} and~\ref{fig:ssouth} as a concentration near the longitude $\mu = -45^\circ$. 
Where non-oblique lattices are included, the concentrations of density along the borders of the QT can be seen, with primitive rectangular lattices appearing as a dense arc on the equator for $ \mu \in [67.5^\circ, 180^\circ)$.

\newcommand{\figurethirteen}{
\begin{figure}
\label{fig:swest}
\caption{
The density map of 2D lattices from CSD crystals on the western hemisphere. 
Angles on the circumference show the latitude $\phi\in[-90^\circ,90^\circ]$.
\textbf{Left}: $N=1100580$ lattices with $\mu\in(-180^\circ,0^\circ]$.
The hexagonal lattice at $\mu = -45^\circ$ and the centered rectangular lattice at $\mu=-112.5^\circ$ are marked on the horizontal arc (western half-equator). 
\textbf{Right}: all $N=932626$ oblique lattices with $\mu\in(-180^\circ,0^\circ]$ and $\phi \neq 0$.}
\includegraphics[width=0.49\textwidth]{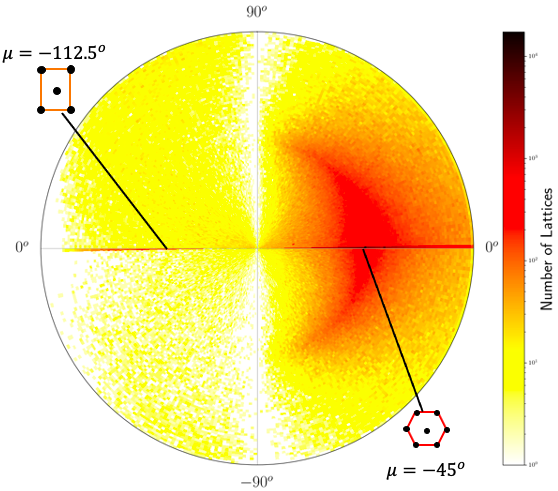}
\includegraphics[width=0.49\textwidth]{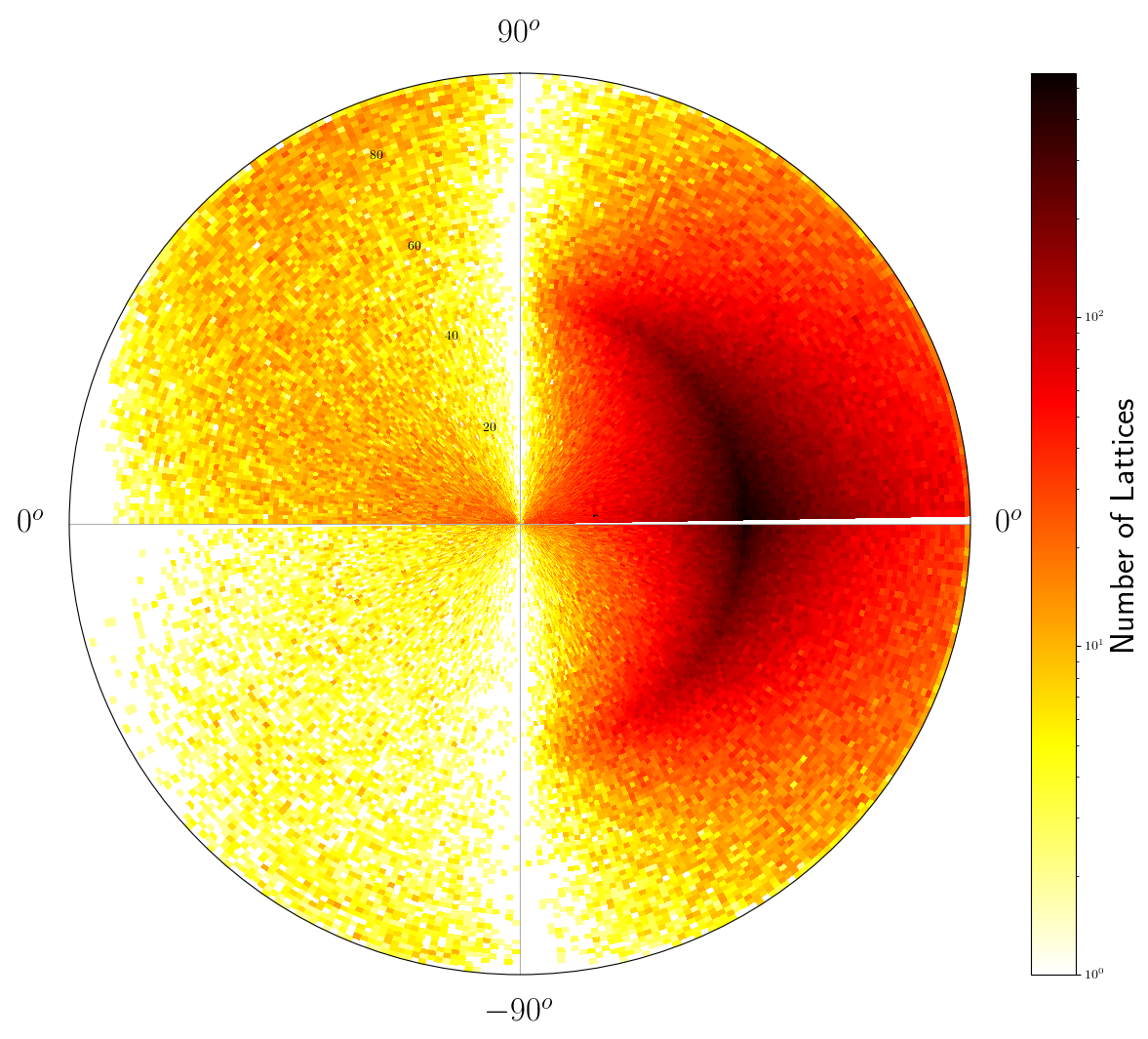}
\end{figure}}
\includefigure{\figurethirteen}

\newcommand{\figurefourteen}{
\begin{figure}
\label{fig:seast}
\caption{
The density map of 2D lattices from CSD crystals on the eastern hemisphere. 
Angles on the circumference show the latitude $\phi\in[-90^\circ,90^\circ]$.
\textbf{Left}: all $N=1511307$ lattices with $\mu\in[0^\circ,180^\circ)$, the square lattice point at $\mu = 67.5^\circ$ and the rectangular lattice at $\mu = 112.5^\circ$ are marked on the the horizontal arc (eastern half-equator). 
\textbf{Right}: all $N=215409$ oblique lattices with $\mu\in [0^\circ,180^\circ)$, $\phi\neq 0$.}
\includegraphics[width=0.49\textwidth]{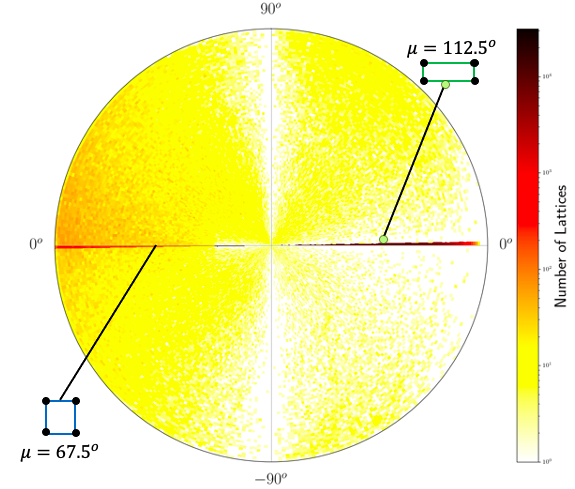}
\includegraphics[width=0.49\textwidth]{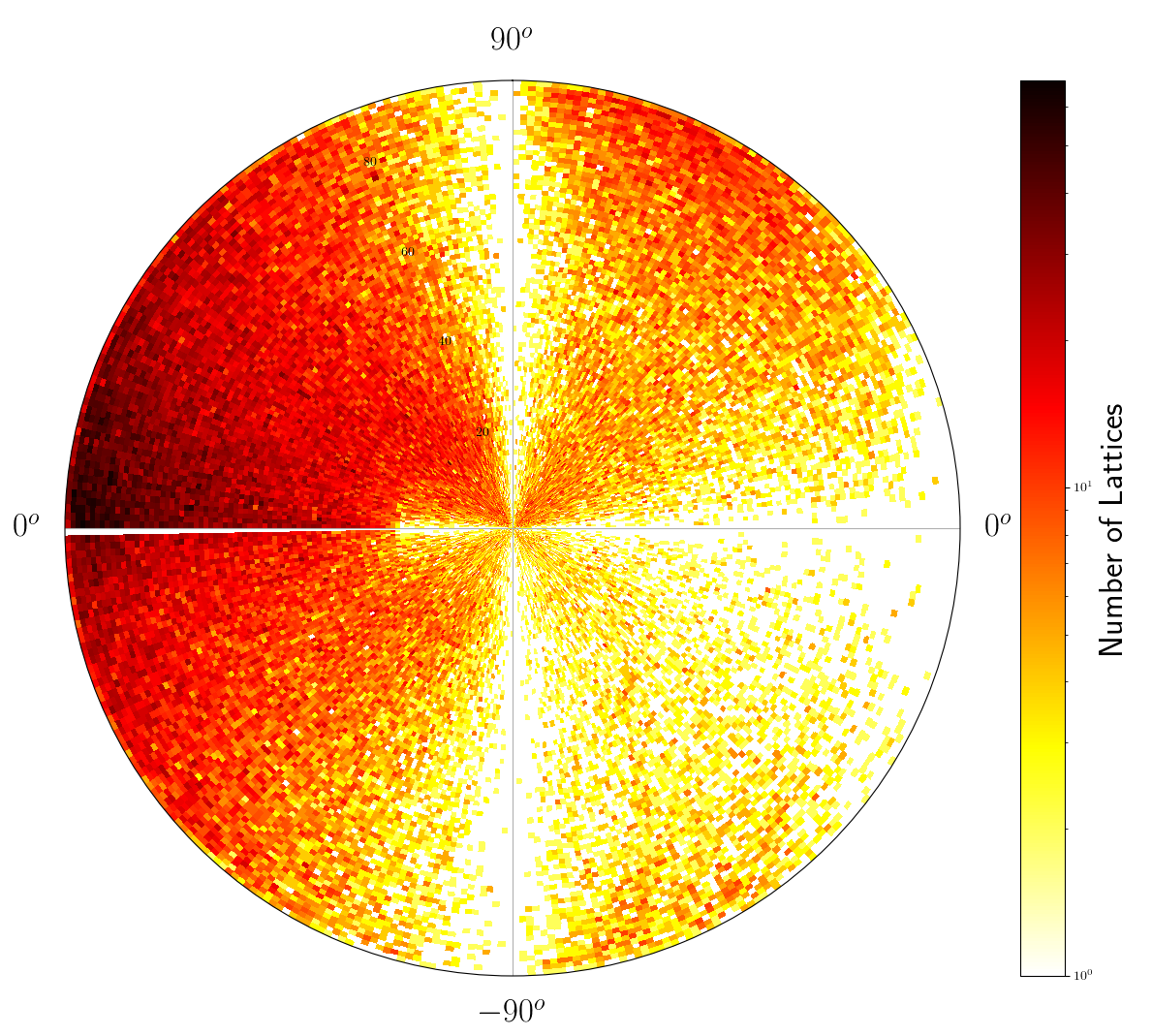}
\end{figure}}
\includefigure{\figurefourteen}

The density maps show a hexagonal `ridge' along the meridional arc at $\mu=-45^\circ$ in Figures~\ref{fig:snorth} and~\ref{fig:ssouth}, which looks as a round arc in Figures~\ref{fig:swest} and~\ref{fig:seast}.
The density of exact square and rectangular lattices is even higher (dark pixels for the Bravais classes tp and op), but there are fewer lattices close to these classes possibly because manual or automatic adjustments are easier for angles close to $90^\circ$ than to $60^\circ$.

\section{Main conclusions and motivations for a continuous crystallography}
\label{sec:conclusions}
 
The density maps in Fig.~\ref{fig:CSD_QS}-\ref{fig:CSD_QS_oblique} and~\ref{fig:snorth}-\ref{fig:seast} visualise for the first time 2.6 million 2-dimensional lattices in real crystals from the Cambridge Structural Database. 
These maps justify the importance of studying latices by continuous invariants and metrics that slightly change under small perturbations, such as the thermal vibrations of atoms.
The Python code is at https://github.com/MattB-242/Lattice\_Invariance. 
\smallskip

Using a geographic analogue, the recent isometry invariants create complete and continuous maps for efficient navigation in the Lattice Isometry Space $\LIS(\R^2)$, which can be zoomed in as satellite images and explored at any desirable resolution.
\smallskip

The four Bravais classes of non-oblique 2D lattices are lower-dimensional subspaces in $\LIS(\R^2)$ whose separate maps in
Fig.~\ref{fig:CSD_rectangular} and~\ref{fig:CSD_hex_square} have no intermediate gaps and contain naturally empty regions only for small or very large values of distance parameters.
\medskip

Using a biological analogue, crystallography previously took a similar approach to the classical taxonomy, dividing lattices into an increasingly complex sequence of discrete categories based on symmetries as they divided organisms according to physical characteristics, see a comprehensive review in \cite{nespolo2018crystallographic}. 
\medskip

The new \emph{continuous crystallography} uses the fundamental geometric properties of the lattice itself to continuously classify an individual in as granular a manner as we like, in a manner akin to the modern use of genetic sequences and markers to classify organisms.  
Indeed, since the root invariant $\RI(\La)$ of a lattice $\La$ is complete, this $\RI(\La)$ could be said to represent the DNA of $\La$.
Even better than the real DNA, any 2D lattice can be explicitly built up from $\RI(\La)$, see \citeasnoun[Proposition~4.9]{kurlin2022mathematics}.

Working towards a materials genome, \citeasnoun[section~7]{widdowson2022average} described complementary invariants, which distinguished all periodic crystals in the CSD and together with invariants of lattices are enough for inverse design of generic crystals.



\ack{Acknowledgements}.
This research was partially supported by the £3.5M EPSRC grant `Application-driven Topological Data Analysis' (2018-2023, EP/R018472/1), the £10M Leverhulme Research Centre for Functional Materials Design (2016-2026) and the last author's Royal Academy of Engineering Fellowship `Data Science for Next Generation Engineering of Solid Crystalline Materials' (2021-2023, IF2122/186).



\referencelist

\vspace*{-15mm}
\appendix

\section{A proof of Proposition~\ref{prop:SM}}
\label{sec:metrics}

\begin{proof}[Proof of Proposition~\ref{prop:SM}]
\textbf{(a)}
For any point $P=(x,y)\in\QT$, the vector $\ora{P^+P}$ has coordinates $(x-t,y-t)$, where $P^+=(t,t)$ is the incentre (the centre of the inscribed circle) of the quotient triangle $\QT$ and $t=1-\dfrac{1}{\sqrt{2}}$, see Fig.~\ref{fig:QT+HS}~(left).
Recall that, for any $b\in\R$, the function $\arctan(b)$ outputs a unique angle $\al\in(-90^\circ,90^\circ)$ such that $\tan(\al)=b$.
If $x>t$, then $\psi=\arctan\dfrac{y-t}{x-t}\in(-90^\circ,90^\circ)$ is the anticlockwise angle from the positive $x$-direction (with the origin at $P^+$) to the vector $\ora{P^+P}$.
\smallskip

For $x=t$, the limit values of arctan give $\psi=\sign(y-t)90^\circ$.
For $x<t$, the anticlockwise angle from the positive $x$-direction to $\ora{P^+P}$ is $\psi+180^\circ$.
For example, the Greenwich vector $\ora{G}$ from the excluded vertex $(1,0)$ to $G=(0,\sqrt{2}-1)\in\QT$ has the anticlockwise angle $\psi+180^\circ=157.5^\circ$
from the positive $x$-direction because 
$$\dfrac{\sqrt{2}-1-t}{-t}=\dfrac{\sqrt{2}-1-(1-\frac{1}{\sqrt{2}})}{\frac{1}{\sqrt{2}}-1}=\dfrac{3-2\sqrt{2}}{1-\sqrt{2}}=1-\sqrt{2}
\text{ and }\arctan(1-\sqrt{2})=-22.5^\circ.$$

\noindent
The anticlockwise angle from the $x$-axis to $\ora{P^+P}$ is
$\al=\left\{\begin{array}{l} 
\psi \text{ if } x>t, \\
\psi+180^\circ \text{ if } x<t, \\
\sign(y-t)90^\circ \text{ if } x=t, y\neq t.
\end{array} \right.$
In all cases above, since the Greenwich vector $\ora{G}$ was chosen as the 0-th meridian, the anticlockwise angle from $\ora{G}$ to $\ora{P^+P}$ is the longitude $\mu=\al-157.5$.
For example, any centred rectangular lattice $\La$ with $\PI(\La)=(x,y)=(\frac{1}{2},\frac{1}{2})$ has $\psi=\arctan\dfrac{y-t}{x-t}=\arctan 1=45^\circ=\al$ and longitude $\mu=\al-157.5^\circ=-112.5^\circ$.
If $\al-157.5^\circ$ is outside the expected range of $\mu\in(-180^\circ,180^\circ]$, we add or subtract $360^\circ$.
Any hexagonal lattice $\La_6$ with $\PI(\La_6)=(0,1)$ has
$\psi=\arctan\dfrac{y-t}{x-t}=\arctan\dfrac{1-(1-\frac{1}{\sqrt{2}})}{\frac{1}{\sqrt{2}}-1}=\arctan\dfrac{1}{1-\sqrt{2}}=-67.5^\circ$, $\al=\psi+180^\circ=112.5^\circ$ and longitude $\mu=\al-157.5^\circ=-45^\circ$. 
Any square lattice $\La_4$ with $\PI(\La_6)=(0,0)$ has
$\psi=\arctan\dfrac{y-t}{x-t}=\arctan 1=45^\circ$, $\al=\psi+180^\circ=225^\circ$ and longitude $\mu=\al-157.5^\circ=67.5^\circ$. 
Formula~(\ref{prop:SM}a) is split into three subcases only to guarantee the range of a longitude $\mu\in(-180^\circ,180^\circ]$ for the anticlockwise angle $\al-157.5^\circ$ from $\ora{G}$ to $\ora{P^+P}$, where $\al$ is computed above.
\smallskip

\noindent
\textbf{(b)}
For a fixed longitude $\mu(\La)$, the projected invariant $\PI(\La)$ varies along the line segment $L$ at a fixed angle from the incentre $P^+$ to the boundary $\bd\QT$.
Formula~(\ref{prop:SM}b) is split into three subcases according to the three boundary edges of $\QT$.
\smallskip

Consider the vertical edge between hexagonal and square lattices, where $\mu(\La)\in[-45^\circ,67.5^\circ]$.
The latitude $\ph(\La)$ is proportional to the ratio in which the point $\PI(\La)=(x,y)$ splits the line segment $L$ from $P^+$ to the vertical edge.
The endpoint $x=0$ means that $\SM(\PI(\La))$ is in the equator with $\ph=0$.
The endpoint $x=t=1-\frac{1}{\sqrt{2}}$ means that $\PI(\La)=P^+$ is in the centre whose image $\SM(P^+)$ is the north pole with $\ph=90^\circ$.
The linear map between these extreme cases gives $\ph(\La)=\dfrac{x}{t}90^\circ=\dfrac{x\sqrt{2}}{\sqrt{2}-1}90^\circ$.
The case of the horizontal edge of $\QT$ gives a similar $\ph$ after replacing $x$ with $y$.
The hypotenuse of $\QT$, where $x+y=1$, is also similar as the incentre $P^+=(x,y)=(1-\frac{1}{\sqrt{2}},1-\frac{1}{\sqrt{2}})$
has the latitude $\ph(\La)=\dfrac{1-x-y}{\sqrt{2}-1}90^\circ=\dfrac{1-2(1-\frac{1}{\sqrt{2}})}{\sqrt{2}-1}90^\circ=90^\circ$ as expected. 
The factor $\sign(\La)$ in~(\ref{prop:SM}b) guarantees a symmetry of $\SM:\QS\to S^2$ in the equator.
\end{proof}

\includeallfigures{
\figureone
\figuretwo
\figurethree
\figurefour
\figurefive
\figuresix
\figureseven
\figureeight
\figurenine
\figureten
\figureeleven
\figuretwelve
\figurethirteen
\figurefourteen
}

\end{document}